\documentclass[11pt]{article}
\pdfoutput=1

\usepackage{hyperref}

\usepackage{graphicx}
\usepackage{epsfig}
\usepackage{here}

\usepackage{amsmath,amssymb} 
\usepackage{mathrsfs} 
\usepackage{mathtools}

\usepackage{amsthm}
\usepackage{braket,mathrsfs,slashed}
\usepackage{ascmac}
\usepackage{fancybox}

\usepackage{multirow}
\usepackage{color}

\usepackage{bbold}
\usepackage{physics}
\usepackage{braket}
\usepackage{slashed}

\usepackage[nosort]{cite}

\usepackage{dsfont}
\usepackage[paper=a4paper,margin=1.5cm]{geometry}
\usepackage[utf8]{inputenc}
\usepackage{rotating}
\usepackage{soul}

\usepackage{multirow}

\usepackage{etoolbox} 

\newlength{\abstractwidth}
\usepackage[fontsize=11.4pt]{scrextend}
\usepackage{setspace}
\usepackage[caption=false]{subfig}

\allowdisplaybreaks

\let\dps\displaystyle
\let\del\partial

\newcommand{\be}{\begin{equation}}
\newcommand{\ee}{\end{equation}}
\renewcommand{\title}[1]{\vbox{\center\bf{\Large{#1}}}\vspace{5mm}}
\renewcommand{\author}[1]{\vbox{\center#1}\vspace{5mm}}
\newcommand{\address}[1]{\vbox{\center\em#1}}

\renewcommand\[{\begin{equation}}
\renewcommand\]{\end{equation}}

\newcommand{\ba}{\begin{eqnarray}}
\newcommand{\ea}{\end{eqnarray}}

\newcommand{\cL}{{\cal L}}
\newcommand{\cO}{{\cal O}}

\DeclareMathOperator{\diag}{\mathrm{diag}} 

\definecolor{crired}{RGB}{153,0,0}    
\definecolor{darkchal}{RGB}{25,25,25}   
\definecolor{charcoal}{RGB}{51,51,51}   
\definecolor{ash}{RGB}{100,100,100}     
\definecolor{pablue}{RGB}{0,102,102}    
\definecolor{tureen}{RGB}{51,153,0}  
\definecolor{paleale}{RGB}{204,204,102}   
\definecolor{lager}{RGB}{140,110,10}    
\definecolor{regal}{RGB}{90,0,120}      
\definecolor{jeans}{RGB}{20,30,150}     

\numberwithin{equation}{section} 
\hypersetup{pdfstartview=FitV,colorlinks=true,linkcolor=midblue,citecolor=midblue,filecolor=midblue,urlcolor=midblue}

\definecolor{midblue}{rgb}{0,0,0.5}

\begin{document}

\newgeometry{top=3.1cm,bottom=3.3cm,right=2.4cm,left=2.4cm}


\begin{titlepage}

\begin{center}
	\hfill \\
	\hfill \\
	\hfill \\
		\vskip 0.5cm

	\title{\Large Topological defects in nonlocal field theories}

	\author{\large
		Luca Buoninfante$^{a,\star}$,
		Yuichi Miyashita$^{b,\dagger}$,
		and Masahide Yamaguchi$^{b,\ddagger}$
	}

	\address{$^a$Nordita, KTH Royal Institute of Technology and Stockholm University,\\
	Hannes Alfvéns väg 12, SE-106 91 Stockholm, Sweden\\
			\vspace{.2cm}
			$^b$Department of Physics, Tokyo Institute of Technology, Tokyo 152-8551, Japan}

\end{center}

\vspace{0.15cm}


\begin{abstract}
	In this paper we study for the first time topological defects in the context of nonlocal field theories in which Lagrangians contain infinite-order differential operators. In particular, we analyze domain walls.
	Despite the complexity of non-linear infinite-order differential equations, we are able to find an approximate analytic solution. 
	We first determine the asymptotic behavior of the nonlocal domain wall close to the vacua. Then, we find a linearized nonlocal solution by perturbing around the well-known local `kink', and show that it is consistent with the asymptotic behavior. 
	We develop a formalism to study the solution around the origin, and use it to verify the validity of the linearized solution.
	We find that nonlocality makes the width of the domain wall thinner, and the energy per unit area smaller as compared to the local case.
	For the specific domain wall solution under investigation we derive a theoretical constraint on the energy scale of nonlocality which must be larger than the corresponding symmetry-breaking scale. We also briefly comment on other topological defects like string and monopole.
\end{abstract}
\vspace{7.7cm}
\noindent\rule{6.5cm}{0.5pt}\\
$\,^\star$
\href{mailto:luca.buoninfante@su.se}{luca.buoninfante@su.se}\\	
$\,^\dagger$ \href{mailto:miyashita.y.ae@m.titech.ac.jp}{miyashita.y.ae@m.titech.ac.jp}\\
$\,^\ddagger$ \href{mailto:gucci@phys.titech.ac.jp}{gucci@phys.titech.ac.jp}

\end{titlepage}

{
	\hypersetup{linkcolor=black}
	\tableofcontents
}

\baselineskip=17.63pt




\newpage

\section{Introduction} \label{intro}

According to the Standard Model of particle physics and to some of its extensions, spontaneous symmetry breaking and phase transitions constitute a crucial aspect for the early evolution of the universe. 

When the temperature starts decreasing because of the expansion of the universe, spontaneous symmetry breaking can be triggered so that the interactions among elementary particles undergo (dis)continuous jumps from one phase to another. New phases with a broken symmetry will form in many regions at the same time, and in each of them only one single vacuum state will be spontaneously chosen. Sufficiently separated spatial regions may not be in causal contact, so that it is quite natural to assume that the early universe is divided into many causally disconnected patches whose size is roughly given by the Hubble radius\footnote{The Hubble radius is given by $R_{\rm H}\sim H^{-1}=a(t)/\dot{a}(t)$ where $a(t)$ is the scale factor depending on the cosmic time $t$. More precisely speaking, the vacuum is chosen over the region with the correlation length of the fields at that time.}, and in each of which the vacuum is independently determined. 
As the universe expands, it can eventually happen that patches with different vacua collide in such a way that boundaries begin to form between adjacent regions with a different vacuum state.
Since the field associated with the spontaneous breaking has to vary in a continuous way between different vacua, it must interpolate smoothly from one vacuum to another via the hill of the potential. This implies that finite-energy field configurations must form at the boundaries separating patches with different vacua, and must persist even after the phase transition is completed. These objects are called \textit{topological defects}~\cite{Vilenkin:2000jqa}, and their formation mechanism (in a cosmological context) is known as \textit{Kibble mechanism}~\cite{Kibble:1976sj,Kibble:1980mv}.

Topological defects can be of several type and different spatial dimensions, and their existence is in one-to-one correspondence with the topology of the vacuum manifold~\cite{Vilenkin:2000jqa}. Domain walls are two-dimensional objects that form when a discrete symmetry is broken, so that the associated vacuum manifold is disconnected. Strings are one-dimensional objects associated to a symmetry breaking whose corresponding vacuum manifold is not simply-connected, and their formation could be predicted both by some extensions of the Standard Model of particle physics and by some classes of Grand Unified Theory (GUT). Monopoles are zero-dimensional objects whose existence is ensured when the vacuum manifold is characterized by non-contractible two-spheres, and they constitute an inevitable prediction of GUT. Moreover, there exist other topological objects called textures that can form when larger groups are broken and whose vacuum-manifold topology is more complicated. 

Since the existence of topological defects is intrinsically related to the particular topology of the vacuum manifold, they can naturally appear in several theories beyond the Standard Model that predict a spontaneous symmetry breaking at some high-energy scale.
For instance, the spontaneous breaking of $SU(5)$ symmetry in GUT leads to the formation of various topological defects.
Therefore, observations and phenomenology of topological defects are very important, and should be considered as unique test-benches to test and constrain theories of particle physics and of the early universe. This also means that for any alternative theory - e.g. that aims at giving a complete ultraviolet (UV) description of the fundamental interactions - it is worth studying the existence of topological defects, investigating how their properties differ with the respect to other models/theories, and putting them to the test with current and future experiments.

In this paper we discuss for the first time topological defects in the context of \textit{nonlocal field theories} in which the Lagrangians contain infinite-order differential operators. In particular, we will make a very detailed analysis of domain wall solutions.
The type of differential operator that we will consider do not lead to ghost degrees of freedom in the particle spectrum despite the presence of higher-order time derivatives in the Lagrangian.

The work is organized as follows:
\begin{description}

\item[\textbf{Sec.~\ref{sec:nonlocal-review}:}] we introduce nonlocal field theories by discussing the underlying motivations and their main properties. 

\item[\textbf{Sec.~\ref{sec-review}:}] we briefly review the domain wall solution in the context of standard (local) two-derivative theories by highlighting various features whose mathematical and physical meanings will be important for the subsequent sections.

\item[\textbf{Sec.~\ref{sec-nft-dw}:}] we analyze for the first time domain wall solutions in the context of ghost-free nonlocal field theories by focusing on the simplest choice for the infinite-order differential operator in the Lagrangian. Despite the high complexity of non-linear and infinite-order differential equations, we will be able to find an approximate analytic solution by relying on the fact that the topological structure of the vacuum manifold ensures the existence of an exact domain wall configuration. 
Firstly, we analytically study the asymptotic behavior of the solution close to the two symmetric vacua. 
Secondly, we find a linearized nonlocal solution by perturbing around the local domain wall configuration. We show that the linearized treatment agrees with the asymptotic analysis, and make remarks on the peculiar behavior close to the origin. We perform an order-of-magnitude estimation of width and energy per unit area of the domain wall. Furthermore, we derive a theoretical lower bound on the scale of nonlocality for the specific domain wall configuration under investigation.

\item[\textbf{Sec.~\ref{sec-other}:}] we briefly comment on other topological defects like string and monopole.

\item[\textbf{Sec.~\ref{sec-dis}:}] we summarize our results, and discuss both theoretical and phenomenological future tasks.

\item[\textbf{App.~\ref{sec-corr}:}] we develop a formalism to confirm the validity of the linearized solution close to the origin.


\item[\textbf{App.~\ref{app-em}:}] we find a compact expression for the canonical energy-momentum tensor in a generic nonlocal (infinite-derivative) field theory.

\end{description}


%
%
We adopt the mostly positive convention for the metric signature, $\eta=\diag(-,+,+,+),$ and work with the natural units system, $c=\hbar=1.$
%
%
%


\section{Nonlocal field theories}\label{sec:nonlocal-review}

The wording `nonlocal theories' is quite generic and, in principle, can refer to very different theories due to the fact that the nonlocal nature of fields can manifest in various ways. In this work with `nonlocality' we specifically mean that Lagrangians are made up of certain non-polynomial differential operators containing infinite-order derivatives.

A generic nonlocal Lagrangian contains both polynomial and non-polynomial differential operators, i.e. given a field $\phi(x)$ one can have
\begin{equation}
\mathcal{L}\equiv\mathcal{L}\left(\phi,\partial\phi,\partial^2\phi,\dots,\partial^n\phi,\frac{1}{\Box}\phi,{\rm ln}\left(- \Box/M_s^2\right)\phi,e^{\Box/M_s^2}\phi,\dots\right),\label{nonlocal-lagr}
\end{equation}
where $\Box=\eta^{\mu\nu}\partial_{\mu}\partial_{\nu}$ is the flat d'Alambertian and $M_s$ is the energy scale at which nonlocal effects are expected to become important.
Non-analytic differential operators like $1/\Box$ and ${\rm log}(-\Box)$ are usually important at infrared (IR) scales, e.g. they can appear as contributions in the finite part of the quantum effective action in standard perturbative quantum field theories~\cite{Barvinsky:2014lja,Woodard:2018gfj}. Whereas analytic operators like $e^{\Box/M_s^2}$ are usually responsible for UV modifications and do not affect the IR physics. Such a transcendental differential operator typically appears in the context of string field theory~\cite{Witten:1985cc,Gross:1987kza,Eliezer:1989cr,Tseytlin:1995uq,Siegel:2003vt,Pius:2016jsl,Erler:2020beb} and p-adic string~\cite{Freund:1987kt,Brekke:1988dg,Freund:1987ck,Frampton:1988kr,Dragovich:2020uwa}.

We are interested in alternative theories that extend the Standard Model in the UV regime, therefore we will focus on analytic differential operators. In general, we can consider a scalar Lagrangian of the following type\footnote{To keep the formula simpler we do not write the scale of nonlocality in the argument of $F(-\Box)$ which, to be more precise, should read $F(-\Box/M_s^2).$}:
\begin{equation}
\mathcal{L}=-\frac{1}{2}\phi F(-\Box)\phi - V(\phi)\,,\label{analytic-Lag}
\end{equation}
where $V(\phi)$ is a potential term, and the kinetic operator can be defined through its Taylor expansion
\begin{equation}
F(-\Box)=\sum\limits_{n=0}^\infty f_n(-\Box)^n\,,\label{nl-oper}
\end{equation}
where $f_n$ are constant coefficients. It should now be clear that the type of nonlocality under investigation manifests through the presence of infinte-order derivatives.

To recover the correct low-energy limit and avoid IR modifications, it is sufficient to require that the function $F(z),$ with $z\in \mathbb{C},$ does not contain any poles in the complex plane. Thus, we choose $F(-\Box)$ to be an \textit{entire function} of the d'Alembertian $\Box$.

By making use of the Weierstrass factorization theorem for entire functions we can write  
\begin{equation}
F(-\Box)=e^{\gamma(-\Box)}\prod\limits_{i=1}^{N}(-\Box+m_i^2)^{r_i}\,,
\end{equation}
where $\gamma(-\Box)$ is another entire function, $m_i^2$ are the zeroes of the kinetic operator $F(-\Box),$ and $r_i$ is the multiplicity of the $i$-th zero. The integer $N\geq 0$ counts the number of zeroes and, in general, can be either finite or infinite. 

To prevent the appearance of ghost degrees of freedom, it is sufficient to exclude the possibility to have extra zeroes besides the standard two-derivative one\footnote{It is worth mentioning that this is \textit{not} the unique possibility for ghost-free higher derivative theories. In fact, we can allow additional pairs of complex conjugate poles and still avoid ghost degrees of freedom and respect unitarity, in both local~\cite{Modesto:2015ozb,Modesto:2016ofr,Anselmi:2017yux,Anselmi:2017lia} and nonlocal theories~\cite{Buoninfante:2018lnh,Buoninfante:2020ctr}. Moreover, tree-level unitarity was shown to be satisfied also if one admits branch cuts in the bare propagator~\cite{Abel:2019ufz,Abel:2019zou}.}. We impose that the kinetic operator does not contain any additional zeroes, so that the effects induced by new physics are entirely captured by the differential operator $e^{\gamma(-\Box)}.$ Therefore, we consider the following Lagrangian:
\begin{equation}
\mathcal{L}=\frac{1}{2}\phi\, e^{\gamma(-\Box)}(\Box-m^2)\,\phi -V(\phi)\,,\label{exp-nonlocal}
\end{equation}
whose propagator reads
\begin{equation}
\Pi(p^2)=-i\frac{e^{-\gamma(p^2)}}{p^2+m^2}\,.\label{propag-nonlocal}
\end{equation}
From the last equation it is evident that no additional pole appears other than $p^2=-m^2,$ because $e^{-\gamma(p^2)}$ is an exponential of an entire function and as such does not have poles in the complex plane.

More generally, under the assumption that the transcendental function $e^{-\gamma(p^2)}$ is convergent in the limits $p^0\rightarrow\pm i\infty,$ it was shown that an S-matrix can be well defined for the Lagrangian~\eqref{exp-nonlocal}, and it can be proven to satisfy perturbative unitarity at any order in loop~\cite{Pius:2016jsl,Briscese:2018oyx,Chin:2018puw,Koshelev:2021orf}. Moreover, the presence of the exponential function can make loop-integrals convergent so that the scalar theory in Eq.~\eqref{exp-nonlocal} turns out to be finite in the high-energy regime~\cite{Krasnikov:1987yj,Moffat:1990jj,Tomboulis:2015gfa,Buoninfante:2018mre}. Very interestingly for such nonlocal theories, despite the presence of infinite-order time derivatives, a consistent initial value problem can be formulated in terms of a finite number of initial conditions~\cite{Barnaby:2007ve,Calcagni:2018lyd,Erbin:2021hkf}.  

This type of transcendental operators with some entire function $\gamma(-\Box)$ have been intensely studied in the past years not only in the context of quantum field theories in flat space~\cite{Krasnikov:1987yj,Moffat:1990jj,Tomboulis:2015gfa,Buoninfante:2018mre,Boos:2018kir,Boos:2019fbu}, but also to formulate ghost-free infinite-derivative theories of gravity~\cite{Krasnikov:1987yj,Kuzmin:1989sp,Tomboulis:1997gg,Biswas:2005qr,Modesto:2011kw,Biswas:2011ar,Frolov:2015bia,Frolov:2015bta,Buoninfante:2018xiw,Buoninfante:2018stt,Buoninfante:2018xif,Koshelev:2016xqb,Koshelev:2017tvv,Koshelev:2020foq,Kolar:2020max}.

In this work we assume that fundamental interactions are intrinsically nonlocal, and that nonlocality becomes relevant in the UV regime. Thus, we consider nonlocal quantum field theories as possible candidates for UV-complete theories beyond the Standard Model. With this in mind, we will analyze topological defects in infinite-derivative field theories, and investigate the physical implications induced by nonlocality in comparison to standard (local) two-derivative theories.

In what follows, we work with the simplest ghost-free nonlocal model for which the entire function is given by
\begin{equation}
\gamma(-\Box)=-\frac{\Box}{M_s^2}\,.\label{entire-box^1}
\end{equation}

\section{Standard domain wall: a brief review} \label{sec-review}

Before discussing domain walls in the context of nonlocal quantum field theories, it is worth reminding some of their basic properties in standard two-derivative field theories, which will then be useful for the main part of this work.

In presence of a domain wall one has to deal with a static scalar field that only depends on one spatial coordinate, e.g. $x,$ and whose Lagrangian reads~\cite{Vilenkin:2000jqa,Saikawa:2017hiv}
\begin{align}
	\cL = \frac{1}{2}\qty(\partial_x\phi)^2 -U(\phi)\,,\qquad U(\phi)=\frac{\lambda}{4}(\phi^2-v^2)^2\,,\label{local-Lag-wall}
\end{align}
which is $\mathbb{Z}_2$-symmetric as it is invariant under the transformation $\phi\rightarrow-\phi;$ $\lambda>0$ is a dimensionless coupling constant and $v>0$ is related to the symmetry-breaking energy scale. The quartic potential has two degenerate minima at $\phi=\pm v$ ($U(\pm v)=0$). 

As mentioned in the Introduction, the discrete symmetry $\mathbb{Z}_2$ can be spontaneously  broken, for instance, in the early universe because of thermal effects. As a consequence, causally disconnected regions of the universe can be characterized by a different choice of the vacuum (i.e. $\phi=+v$ or $\phi=-v$), and when two regions with different vacua collide a continuous two-dimensional object -- called domain wall -- must form at the boundary of these two regions.

Let us now determine explicitly such a finite-energy configuration interpolating $\pm v.$
First of all, we impose the asymptotic boundary conditions
\begin{align}
	\phi(-\infty) = -v\,,
	\qquad
	\phi(\infty) = v\,.\label{boundary conditions}
\end{align}
The field configuration must be non-singular and of finite energy, therefore $\phi(x)$ must interpolate smoothly between the two vacua, this implies that there exists a point $x_0\in \mathbb{R}$ such that $\phi(x_0)=0.$ Without any loss of generality, we can choose the reference frame such that the centre of the wall is at the origin $x_0=0,$ i.e. $\phi(0)=0.$
%

The energy density can be computed as
\begin{align}
	\mathcal{E}(x) \equiv T_{0}^0(x)= \frac{1}{2}(\del_x\phi)^2 + \frac{\lambda}{4}(\phi^2-v^2)^2\,,\label{local-e-dens}
\end{align}
from which it follows  $\mathcal{E}(x)\geq U(0)=\lambda v^2/4,$ and this implies that  there exists a solution that does not dissipate at infinity. Hence, the topological structure of the vacuum manifold -- which is disconnected in the case of $\mathbb{Z}_2$ symmetry -- ensures the existence of a non-trivial field configuration of finite energy. 

We can determine qualitatively the behavior of this field configuration by making an order-of-magnitude estimation of the width $R$ (along the $x$-direction), and of the energy per unit area $E$ of the wall.

In fact, we can define the width of the wall in three ways. The first one is to use the energy density in Eq.~\eqref{local-e-dens}. The lowest energy configuration interpolating the two vacua can be found by balancing the kinetic and potential term in the energy density $\mathcal{E}(x)$.

By approximating the gradient with the inverse of the width, $\partial_x\sim 1/R,$ and the field value with $\phi\sim v,$ Eq.~\eqref{local-e-dens} gives
\begin{align}
	\frac{1}{2} \frac{1}{R^2} v^2 \sim \frac{\lambda}{4} v^4\quad
		\Rightarrow\quad
	R \sim \sqrt{\frac{2}{\lambda}} \frac{1}{v}\,,\label{local-radius-estim}
\end{align}
from which it follows that the width of the wall is of the same order of the Compton wavelength $R\sim(\sqrt{\lambda}v)^{-1}\sim m^{-1}$.
Whereas, the energy per unit area can be estimated as
\begin{eqnarray}
	E&=& \int_\mathbb{R}{{\rm d}x}\qty[\frac{1}{2}(\del_x\phi)^2+\frac{\lambda}{4}(\phi^2-v^2)^2] \nonumber\\[2mm]
	 &\sim& (\text{width of the wall}) \times (\text{energy density}) \nonumber\\[2mm]
		& \sim& R \times \lambda v^4 \nonumber\\[2mm]
		&\sim&\sqrt{\lambda}v^3\,.\label{local-energ-estim}
\end{eqnarray}
The other two ways are to use the exact configuration (solution) of the domain wall. 
The field equation 
\begin{align}
	\partial_x^2 \phi(x) = \lambda \phi(\phi^2-v^2)\label{local-field-eq}
\end{align}
can be solved by quadrature, and an exact analytic solution can be found, and it satisfies all the qualitative properties discussed above. 
The exact solution is sometime called `kink', and it reads~\cite{Vilenkin:2000jqa}
\begin{align}
	\phi(x) = v \tanh \left(\sqrt{\frac{\lambda}{2}}vx\right)\,.\label{local-dom-wal-sol}
\end{align}
Its asymptotic behavior is given by
\begin{align}
|x|\rightarrow \infty\quad \Rightarrow \quad \phi(x)\sim \pm v\qty(1-2e^{-\sqrt{2\lambda}vx})\,. 
		\label{eq:asymp sol in local case}
\end{align}
Through the exact solution~\eqref{local-dom-wal-sol}, we can define the width of the wall in two ways. One way is to identify it with the typical length scale over which $\phi(x)$ changes in proximity of the origin, that is, the length scale $\ell$ defined as the inverse of the gradient at the origin, i.e. $\ell\sim v/(\partial_x\phi|_{x=0}),$ where the scale $v$ is introduced for dimensional reasons. From Eq.~\eqref{local-dom-wal-sol} we have
\begin{equation}
\del_x\phi(x)|_{x=0}=v^2\sqrt{\frac{\lambda}{2}},
\end{equation}
which yields
\begin{equation}
\ell\sim \sqrt{\frac{2}{\lambda}}\frac{1}{v} = R\,.\label{ell-linearized-local}
\end{equation}
The last way is to use the asymptotic behavior given in Eq.~\eqref{eq:asymp sol in local case}. The width of the wall, $\widetilde{R}$, can be defined as 
\begin{align}
|x|\rightarrow \infty\quad \Rightarrow \quad \phi(x)\sim \pm v\qty(1-2e^{-\frac{2x}{\widetilde{R}}})\,, 
		\label{eq:asymp radius}
\end{align}
which yields $\widetilde{R} \sim \sqrt{\frac{2}{\lambda}} \frac{1}{v} = R = \ell\,.$ In the local case, all of the three definitions give the same expressions and we need not discriminate them. But, as we will show, in the nonlocal case, all of the definitions would give different expressions in the sub-leading order, though two of them ($R$ and $\widetilde{R}$) have the similar feature. In fact, $R$ and/or $\widetilde{R}$ might be more appropriate as the definition of the width (or the radius) of a domain wall because $\ell$ is related to the behavior of the solution close to the origin and far from the vacuum.

We can obtain the energy per unit area as 
\begin{eqnarray}
       E = \int_{\mathbb{R}} {\rm d}x\mathcal{E}(x) =\int_{\mathbb{R}} {\rm d}x\qty(\frac{{\rm d}\phi}{{\rm d}x})^2
		= \frac{4}{3} \sqrt{\frac{\lambda}{2}}v^3\,,
\end{eqnarray}
which is consistent with the estimation in Eq.~\eqref{local-energ-estim} up to an order-one numerical factor.

The discussion in this Section was performed for a local two-derivative theory in one spatial dimension. However, the essential concepts and methods, like the condition for the existence of a solution related to the non-trivial topology of the vacuum manifold, and the order-of-magnitude estimations, can be applied to nonlocal field theories and to higher dimensional cases (e.g. string and monopole).


\section{Domain wall in nonlocal field theories} \label{sec-nft-dw}

In this Section we analyze the domain wall solution for the nonlocal field theory~\eqref{exp-nonlocal} with the simplest choice of entire function given in~\eqref{entire-box^1}. Hence, we consider a nonlocal generalization of the Lagrangian~\eqref{eq:Lagrangian for DW} given by
\begin{align}
	\cL
		= \frac{1}{2}\phi e^{-\partial_x^2/M_s^2}(\partial_x^2+\lambda v^2)\phi
				-\frac{\lambda}{4}\qty(\phi^4+v^4)\,,
	\label{eq:Lagrangian for DW}
\end{align}
whose field equation reads
\begin{align}
	e^{-\partial_x^2/M_s^2}(\partial_x^2+\lambda v^2)\phi = \lambda\phi^3\,.
		\label{eq:the equation of the motion for DW}
\end{align}
We can easily verify that in the limit $\partial_x^2/M_s^2\rightarrow 0$ we consistently recover the local two-derivative case, i.e. Eqs.~\eqref{local-Lag-wall} and~\eqref{local-field-eq}.

In this case the differential equation is non-linear and highly nonlocal, and finding a solution seems to be very difficult not only analytically but even numerically. However, despite the complexity of the scalar field equation, we can still find a domain wall configuration. 

First of all, by relying on the fact that the presence of the exponential operator $e^{-\partial_x^2/M_s^2}$ should not change the number of degrees of freedom and of initial conditions~\cite{Barnaby:2007ve}, we can impose the same boundary conditions as done in the two-derivative case in Eq.~\eqref{boundary conditions}, i.e. $\phi(\pm \infty)=\pm v.$ 

Furthermore, the existence of a time-independent and non-dissipative solution is still guaranteed by the non-trivial topological structure of the vacuum manifold which is disconnected in the case of $\mathbb{Z}_2$ symmetry. In other words, also in the nonlocal case a finite-energy field configuration that smoothly interpolates between the two vacua $\phi=\pm v$ must exist. Indeed, the energy density can be computed from the expression of the energy-momentum  tensor~\eqref{stress-final} (with $m^2=-\lambda v^2$) and, up to total derivatives, we have: 
\begin{align}
	\mathcal{E}(x)
		\equiv T_0^0(x)= -\frac{1}{2}\phi e^{-\partial_x^2/M_s^2}(\partial_x^2+\lambda v^2)\phi
			+\frac{\lambda}{4}(\phi^4+v^4)\,,
	\label{eq:energy density for DW}
\end{align}
which can be easily verified to be positive definite after imposing the field equation and using the fact that the solution satisfies $|\phi|\leq v$. Since $\phi(v)=-\phi(-v)$ and by continuity in $x,$ there must exist a point $x_0$ such that $\phi(x_0)=0.$ By translation invariance we can set $x_0=0$ and $\phi(0)=0,$ without any loss of generality. It follows that $\mathcal{E}(x)\geq U(0)=\lambda v^2/4$ for any $x\in \mathbb{R},$ which implies the existence of a time-independent solution that does not dissipate at infinity. 

Hence, also for the nonlocal model under investigation the topological structure of the vacuum manifold ensures the existence of a non-trivial field configuration of finite energy which does not dissipate at infinity. Topology also implies
that the solution is stable against time-independent perturbations of its spatial shape. 

%


\begin{figure}[t]
	\centering
	\includegraphics[scale=0.6]{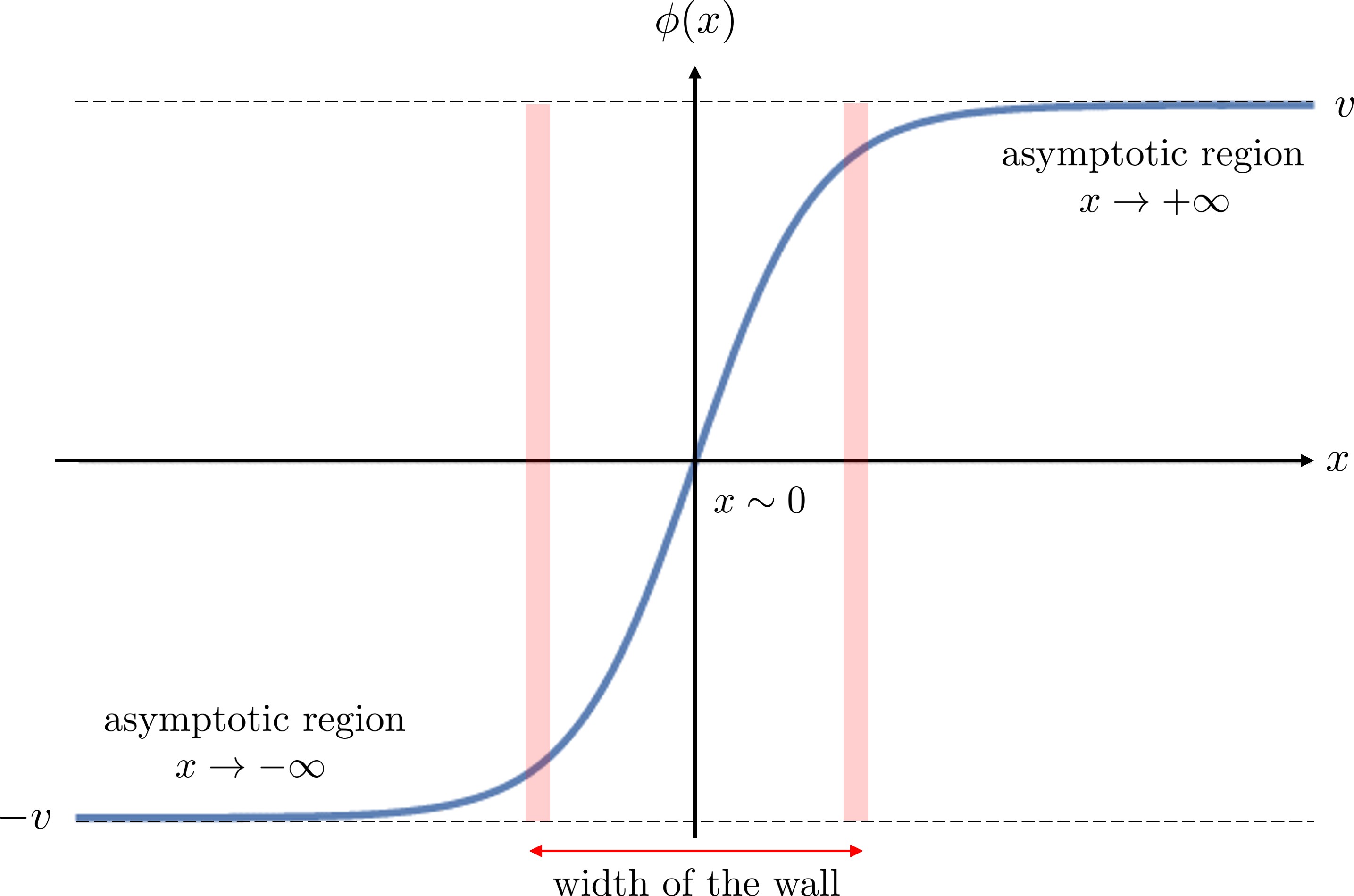}
	\caption{Schematic illustration of the analysis made in this Section to study the domain wall configuration in nonlocal field theory, whose qualitative behavior is drawn consistently with the boundary conditions $\phi(\pm\infty)=\pm v$, and with the choice of the origin $\phi(0)=0.$ We analyze the behavior of the domain wall configuration in several regimes, and use different methods to find approximate analytic solutions. (i) We study the asymptotic behavior of the domain wall solution in the limit $|x|\rightarrow \infty.$ (ii) We find a linearized nonlocal solution by perturbing around the local domain wall configuration treated as a background, and analyse its behavior not only at infinity but also close to the origin. (iii) We make an order-of-magnitude estimation for the width and the energy per unit area of the wall, and verify the consistency with the analytic approximate solutions.}
	\label{fig:conceptual figure for analysis of nonlocal domain wall}
\end{figure}
%


Very interestingly, knowing that a domain wall solution must exist, and equipped with a set of boundary conditions, we can still find an approximate analytic solution by working perturbatively in some regime. We will proceed as follows. In Sec.~\ref{subsec-asym} we will determine the solution asymptotically close to the vacua $\pm v$ (i.e. for $|x|\rightarrow \infty$). In Sec.~\ref{subsec-pert} we will find a linearized nonlocal solution by perturbing around the known local `kink' configuration treated as a background. Finally, in Sec.~\ref{subsec-order} we will make an order-of-magnitude estimation for the width and the energy of the nonlocal domain wall, and check the consistency with the approximate analytic solutions. See Fig.~\ref{fig:conceptual figure for analysis of nonlocal domain wall} for a schematic illustration of our analysis.

\paragraph{Remark 1.} The presence of the infinite-derivative differential operator in the Lagrangian requires some more discussion. Because of the minus sign in the exponent it is not always guaranteed that its action on any function is well-defined. For example, given a function $f(x)$ that admits Fourier transform, a term like $e^{-\partial_x^2/M_s^2}f(x)$ in Fourier space becomes $e^{+k^2/M_s^2}\tilde{f}(k).$ Thus, we should make sure to work with a class of functions on which the action of the nonlocal operator gives a finite result despite the plus sign in the exponent of $e^{+k^2/M_s^2}$. In this paper, we implicitly work with a restricted class of functions (either Fourier transformable or not) such that one can still produce a finite result after acting with $e^{-\partial_x^2/M_s^2}$. It is worth to mention that the same issue was discussed in the context of string field theory~\cite{Witten:1985cc}. 
Below we will confirm that the action of the nonlocal operator in this work is well-defined;  for instance, for the asymptotic analysis in Sec.~\ref{subsec-asym} we will have $f(x)= e^{-Bx}$ on which we can safely define the action of the infinite-derivative operator.

\subsection{Asymptotic solution for $|x|\rightarrow \infty$} \label{subsec-asym}

As a first step we analyze the asymptotic behavior of the solution close to the two vacua $\phi=\pm v$, i.e. in the regime $|x|\rightarrow \infty.$ Let us first consider the perturbation around $\phi=+v,$ i.e. we write 
\begin{equation}
    \phi(x)=v+\delta\phi(x)\,,\qquad \frac{|\delta\phi|}{v}\ll 1\,,
\end{equation}
so that the linearized field equation reads
\begin{align}
	e^{-\del_x^2/M_s^2} (\del_x^2+\lambda v^2) \delta\phi = 3\lambda v^2 \delta\phi\,.
\end{align}
Taking inspiration from the asymptotic behavior of the domain wall in the local case (see Eq.~\eqref{eq:asymp sol in local case}), as an ansatz we assume that $\phi(x)$ approaches the vacuum exponentially, i.e. we take 
\begin{align}
	\delta\phi = \phi - v = A e^{-B x} \label{eq:def for asymp sol}
\end{align}
where $A,$ and $B>0$ are two constants. 

Since the exponential is an eigenfunction of the kinetic operator, we can easily obtain an equation for $B,$
\begin{align}
	e^{-B^2/M_s^2} (B^2+\lambda v^2)  = 3\lambda v^2\,.\label{eq-for-B}
\end{align}
By using the principal branch $W_0(x)$ of the Lambert-W function (defined as the inverse function of $f(x)=xe^x$) we can solve Eq.~\eqref{eq-for-B} as follows
\begin{align}
	B^2 = - M_s^2\, W_0\qty(-\frac{3\lambda v^2}{M_s^2}e^{-\lambda v^2/M_s^2})-\lambda v^2\,.
		\label{eq:coeff B in asymp sol}
\end{align}
Before continuing let us make some remarks on the Lambert-W function. It is a multivalued function that has an infinite number of branches $W_n$ with $n\in \mathbb{Z}.$ The only real solutions are given by the branch $W_0(x)$ for $x\geq -1/e,$ and an additional real solution comes from the branch $W_{-1}(x)$ for $-1/e\leq x<0.$ In the equations above we have taken the so-called principal branch $W_0,$ and we will do the same in the rest of the paper. However, we will also comment on the branch $W_{-1}$ and the physical implications associated to it. Regarding higher order branches $n>0,$ they will generate non-physical complex values, and in some cases they do not even recover the local limit; therefore, we discard such solutions.

Note that, by means of the asymptotic analysis above the coefficient $A$ cannot be determined as it factors out from the field equation but, as we will explain in Sec.~\ref{subsec-pert}, we will be able to determine it up to order $\mathcal{O}(1/M_s^2)$. 


\subsubsection{A theoretical constraint on the scale of nonlocality} \label{subsubsec-cond}

The use of the Lambert-W function to obtain the solution for $B$ in Eq.~\eqref{eq:coeff B in asymp sol} relied on the fact that  Eq.~\eqref{eq-for-B} could be inverted. As explained above this inversion is valid if and only if $c \in \{xe^x|x\in\mathbb{R}\}$, i.e. if $c\geq -1/e$.

Applying this condition to~\eqref{eq:coeff B in asymp sol}, we get
\begin{align}
	-\frac{3\lambda v^2}{M_s^2}e^{-\lambda v^2/M_s^2}\geq -\frac{1}{e}\,,\label{condition}
\end{align}
and inverting in terms of the (principal branch) Lambert-W function we obtain the following \textit{theoretical constraint}:
\begin{align}
M_s^2\geq  -\frac{\lambda v^2}{W_0\qty(-1/3e)}\,,	\label{eq:neccessary cond for asymp sol}
\end{align}
where we have used the fact that $W_0(x)$ is a monotonically increasing function. The inequality~\eqref{eq:neccessary cond for asymp sol} means that 
the energy scale of nonlocality must be greater than the symmetry-breaking scale $\sqrt{\lambda} v.$ We can evaluate $W_0\qty(-1/3e)\simeq -0.14,$  so that the lower bound reads $M_s^2\gtrsim 7.14 \lambda v^2.$
One usually obtains constraints on the free parameters of a theory by using experimental data. In the present work, instead, we found a purely theoretical constraint. See Sec.~\ref{sec-dis} for further discussions on this feature.

Given the fact that $\lambda v^2/M_s^2< 0.14,$ we can expand~\eqref{eq:coeff B in asymp sol} and obtain
\begin{align}
	B^2
		= 2 \lambda v^2 \qty(1+\frac{3\lambda v^2}{M_s^2})
			+ \cO\qty(\qty(\frac{\lambda v^2}{M_s^2})^2)\,,
	\label{eq:series expansion for coeff B}
\end{align}
or by taking the square root,
\begin{align}
	B
		= \sqrt{2 \lambda} v \qty(1+\frac{3}{2}\frac{\lambda v^2}{M_s^2})
			+ \cO\qty(\qty(\frac{\lambda v^2}{M_s^2})^2)\,\,.
	\label{eq:series expansion for coeff B-sqrt}
\end{align}
From this expression, we can obtain the width of wall, $\widetilde{R}$, defined in Eq.~\eqref{eq:asymp radius} as
\begin{equation}
\widetilde{R} \sim \frac{2}{B}\sim \sqrt{\frac{2}{\lambda}}\frac{1}{v}\left(1-\frac32\frac{\lambda v^2}{M_s^2}\right)\,.\label{eq:tildeR}
\end{equation}
Also, from Eq.~\eqref{eq:series expansion for coeff B-sqrt} we can check that in the local limit $M_s\rightarrow \infty$ we recover the two-derivative case in Eq.~\eqref{eq:asymp sol in local case}:
\begin{align}
	\lim_{M_s\to\infty} B^2 = 2 \lambda v^2 = \qty(\sqrt{2\lambda}v)^2\equiv B_{\rm L}^2\,.
\end{align}
Furthermore, from~\eqref{eq:series expansion for coeff B} we can notice that $B\geq B_{\rm L}=\qty(\sqrt{2\lambda}v).$ This physically means that the nonlocal domain wall solution approaches the vacuum $\phi=+v$ \textit{faster} as compared to the local two-derivative case. This feature, which is manifest in Eq.~\eqref{eq:tildeR}, may also suggest that the width of the nonlocal domain wall is \textit{smaller} as compared to the local case; indeed this fact will also be observed with the expression of $R$ in the next Subsections.

So far we have only focused on the asymptotic solution for $x\rightarrow +\infty$ ($\phi(+\infty)=+v$) but the same analysis can be applied to the other asymptotic $x\rightarrow -\infty$ ($\phi(-\infty)=-v$), and the same results hold because of the $\mathbb{Z}_2$ symmetry.

\paragraph{Remark 2.} Before concluding this Subsection it is worth commenting on the validity of the asymptotic solution we determined. On one hand, we know that the existence of the domain wall is ensured by the topological structure of the vacuum manifold, and this should not depend on the value of $M_s.$ On the other hand, it appears that the asymptotic solution we found is only valid for some values of $M_s$ satisfying the inequality in Eq.~\eqref{eq:neccessary cond for asymp sol}, which seems to imply that the domain wall solution does not exist for other values of $M_s$. Is this a contradiction? The answer is no, all is consistent, and in fact the domain wall solution exists for any value of $M_s.$ 

First of all, we should note that to solve Eq.~\eqref{condition} in terms of $M_s$ we have used the principal branch $W_0(x)$ which is a monotonic increasing function, but an additional real solution can be found by using the branch $W_{-1}(x)$ which is, instead, a monotonically decreasing function. Thus, given the opposite monotonicity behavior of $W_{-1}$ as compared to $W_{0},$ if we solve Eq.~\eqref{condition} by means of $W_{-1}$ we get $M_s\leq -\lambda v^2/W_{-1}(-1/3e)\simeq 0.30\lambda v^2.$ 
Moreover, in the range of values $0.30 \lambda v^2 \lesssim M_s^2\lesssim 7.21 \lambda v^2$ the functional form in Eq.~\eqref{eq:def for asymp sol} does not represent a valid asymptotic behavior for the domain wall. In this case the domain wall configuration may be characterized by a completely different profile, but its existence is still guaranteed by the non-trivial topology.
Anyway, as already mentioned above, in this paper we only work with $W_0,$ therefore with values of $M_s$ satisfying the inequality~\eqref{eq:neccessary cond for asymp sol}. See also Sec.~\ref{sec-dis} for more discussion on this in relation to physical implications.


\subsection{Perturbation around the local solution} \label{subsec-pert}

Let us now implement an alternative method to determine the behavior of the nonlocal domain wall not only at infinity but also close to the origin. 

We consider a linear perturbation around the standard two-derivative domain wall configuration $\phi_{\rm L}(x)=v\tanh(\sqrt{\lambda/2} vx)$. Let us define the deviation from the local solution as 
\begin{align}
	\delta\phi (x) = \phi(x) - \phi_{\rm L}(x)\,, \qquad  \left|\frac{\delta\phi}{\phi_{\rm L}}\right|\ll 1\,, \label{pert-loc-nonloc}
\end{align}
in terms of which we can linearize the field equation~\eqref{eq:the equation of the motion for DW}:
\begin{eqnarray}
\left[e^{-\partial_x^2/M_s^2}(\partial_x^2+\lambda v^2)-3\lambda\phi_{\rm L}^2\right]\delta\phi=\lambda (1-e^{-\partial_x^2/M_s^2})\phi_{\rm L}^3\,.\label{linear-lnl}
\end{eqnarray}
Since the nonlocal scale appears squared in~\eqref{eq:the equation of the motion for DW}, we would expect that $\delta\phi\sim \mathcal{O}(1/M_s^2)$ such that the local limit is consistently recovered, i.e. $\delta\phi\to0$ when $M_s^2\to\infty$. We now write Eq.~\eqref{pert-loc-nonloc} up to order $\mathcal{O}(1/M_s^2)$ in order to extract the leading nonlocal correction to $\phi_{\rm L}.$ By expanding the nonlocal terms as follows  
\begin{eqnarray}
	e^{-\del_x^2/M_s^2} \delta\phi&=&
	\delta\phi 
	+ \cO\qty(\frac{\del_x^2}{M_s^2}) \delta\phi\,,\\[2mm]
	e^{-\del_x^2/M_s^2} \phi_{\rm L}^3&=& \phi_{\rm L}^3-\frac{\partial_x^2}{M_s^2}\phi_{\rm L}^3
	+\cO\qty(\qty(\frac{\del_x^2}{M_s^2})^2)\phi_{\rm L}^3\,,
\end{eqnarray}
we can write~\eqref{linear-lnl} up to order $\mathcal{O}(1/M_s^2):$
\begin{align}
	\qty[\partial_x^2+\lambda v^2-3\lambda\phi_{\rm L}^2(x)] \delta\phi
		= \frac{\lambda}{M_s^2}\del_x^2(\phi_{\rm L}(x)^3)\,.
		\label{lin-eq-nln-2}
\end{align}
This expansion is valid as long as the following inequality holds:
\begin{equation}
\frac{1}{|\delta\phi|}\left|\frac{\del_x^2}{M_s^2}\delta\phi\right| \ll 1\,.\label{ineq-2}
\end{equation}

We now introduce the dimensionless variable $s=\sqrt{\lambda/2}\,vx$ and the function $f(s)=\delta\phi(x)/v,$ so that we can recast Eq.~\eqref{lin-eq-nln-2} as
\begin{align}
	f^{\prime\prime}(s)+2(1-3\tanh^2 s)f(s) = \frac{\lambda v^2}{M_s^2}(\tanh^3 s)^{\prime\prime}\label{diff-eq-s}
\end{align}
where the prime $'$ denotes the derivative with respect to $s$. The above differential equation can be solved analytically, and its solution reads
\begin{eqnarray}
\!\!f(s)&=&
	\frac{C_1}{\cosh^2s}
	+ \frac{3C_2/2+27\lambda v^2/32M_s^2}{\cosh^2s} \log\frac{1+\tanh s}{1-\tanh s}
\nonumber\\[2mm]
	&&+\qty[
		\qty(2C_2+\frac{\lambda v^2}{8M_s^2})\cosh^2s
			+ \qty(3C_2-\frac{61\lambda v^2}{16M_s^2})
			+ \frac{2\lambda v^2}{M_s^2}(1+\tanh^2 s)
	] \tanh s\,,\,\,\,
\end{eqnarray}
where $C_1$ and $C_2$ are two integration constants to be determined.

\begin{figure}[t]
	\centering
	\includegraphics[scale=0.48]{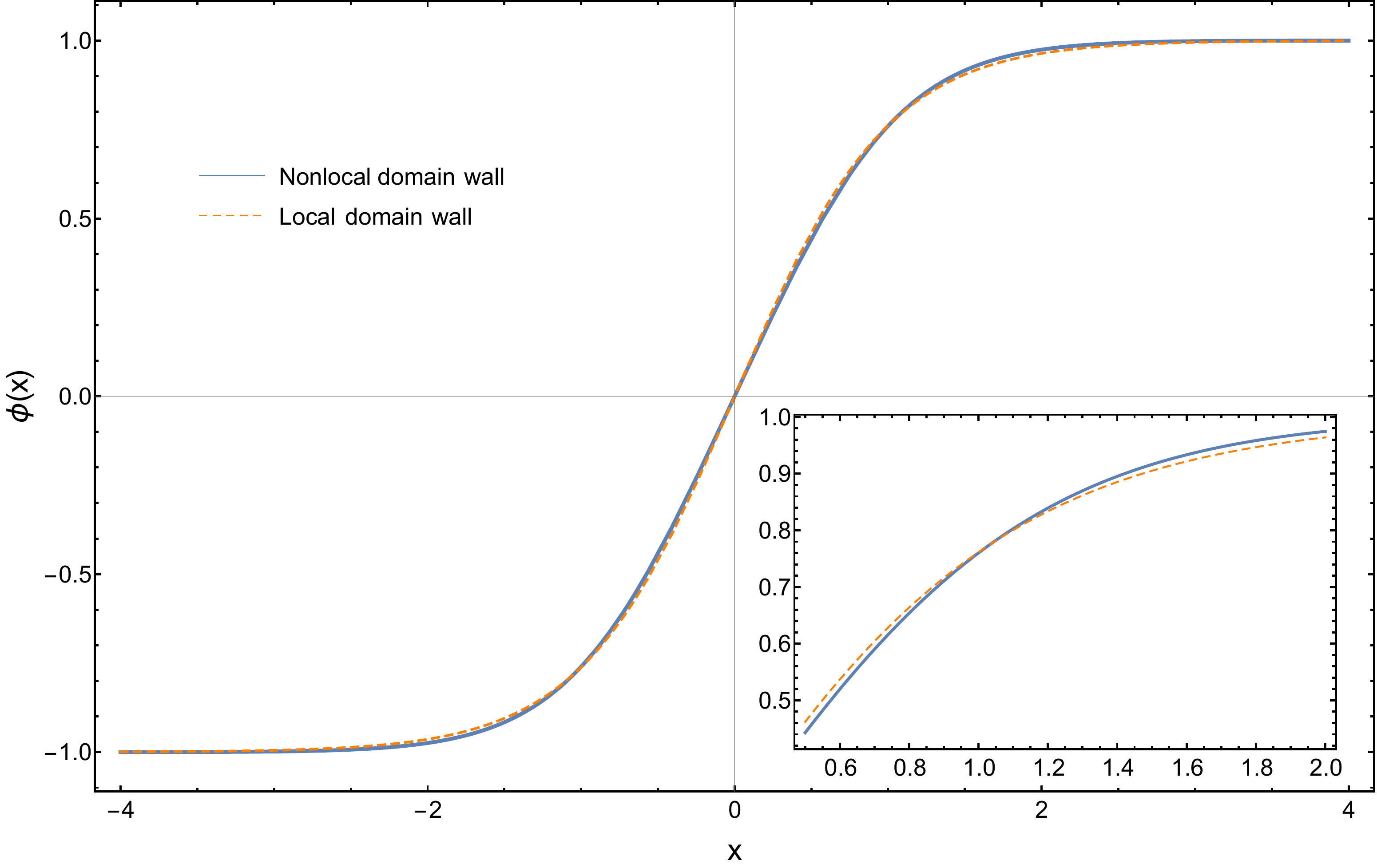}
	\caption{
			In this figure we show the linearized nonlocal domain wall solution $\phi(x)=\phi_{\rm L}(x)+\delta\phi(x)$ (solid blue line) in comparison with the local domain wall (orange dashed line). The nonlocal configuration approaches the asymptotic vacua at $x\rightarrow \pm \infty$ faster as compared to the local case.
			In the smaller plot we showed the behavior of the two solutions over a smaller interval in order to make more evident the differences between local and nonlocal cases. We can notice that when going from $x=0$ to $x\rightarrow \pm \infty$ the nonlocal curve slightly oscillates around the local one.
			We set $\lambda=2$, $v=1$ and $M_s^2=14.3,$ which are consistent with the theoretical constraint $M_s^2\geq -\lambda v^2/W_0(-1/3e)$ in Eq.~\eqref{eq:neccessary cond for asymp sol}.
			}
	\label{fig:illustration for the perturbative solution}
\end{figure}

The boundary conditions $\phi(\pm \infty)=\pm v$ in terms of the linearized deviation read $\delta\phi(\pm \infty)=0,$ or equivalently $f(\pm \infty)=0$. These are satisfied if and only if the algebraic relation $2C_2+\lambda v^2/8M_s^2=0$ holds true, which means that $C_2=-\lambda v^2/16M_s^2$.
Moreover, the constant $C_1$ must be zero because of the $\mathbb{Z}_2$-symmetry.
Thus, the solution for $\delta \phi$ is given by
\begin{align}
	\delta\phi
		= vf(s)
		= \frac{\lambda v^3}{M_s^2}\frac{1}{\cosh^2\sqrt{\frac{\lambda}{2}}vx}
			\qty[\frac{3}{4}\log\frac{1+\tanh\sqrt{\frac{\lambda}{2}}vx}{1-\tanh\sqrt{\frac{\lambda}{2}}vx}
					-2\tanh\sqrt{\frac{\lambda}{2}}vx]\,.
	\label{eq:exact sol for delta phi}
\end{align}
In Fig.~\ref{fig:illustration for the perturbative solution} we showed the behavior of the nonlocal domain wall solution $\phi=\phi_{\rm L}+\delta\phi$ in comparison with the local two-derivative one $\phi_{\rm L};$ we have set values for $v,$ $\lambda$ and $M_s$ consistently with the theoretical lower bound in Eq.~\eqref{eq:neccessary cond for asymp sol}. From the plot we can notice that the nonlocal solution approaches the vacua $\pm v$ faster as compared to the local case, which is in agreement with the asymptotic analysis in the previous Subsection. Indeed, we can expand the solution $\phi=\phi_{\rm L}+\delta\phi$ in the regime $|x|\rightarrow \infty$, and obtain
\begin{eqnarray}
\phi(x)&\simeq& \pm v\left[1-2\left(\frac{4v^2\lambda}{M^2_s}\right)e^{-\sqrt{2\lambda}vx}-2e^{-\sqrt{2\lambda}vx}\left(1-\frac{3}{2}\sqrt{2\lambda}\frac{\lambda v^3}{M_s^2}x\right)\right]\nonumber\\[2mm]
&\simeq& \pm v\left[1-2\left(1+\frac{4v^2\lambda}{M^2_s}\right)e^{-\sqrt{2\lambda}v(1+3\lambda v^2/2M_s^2)x}\right]\nonumber\\[2mm]
&\simeq& \pm v\left[1-2\left(1+\frac{4v^2\lambda}{M^2_s}\right)e^{-B x}\right]\,,
\end{eqnarray}
where to go from the first to the second line we have used the freedom to add negligible terms of order higher than $\mathcal{O}(1/M_s^2),$ i.e. $4v^2\lambda/M_s^2\simeq 4v^2\lambda/M_s^2(1-3\sqrt{2\lambda}\lambda v^3x/2M_s^2)$ and $1-3\sqrt{2\lambda}\lambda v^3/2M_s^2x\simeq e^{-(3\sqrt{2\lambda}\lambda v^3/2M_s^2)x}.$ Remarkably, the asymptotic behavior of the linearized solution perfectly matches the result obtained in Eq.~\eqref{eq:series expansion for coeff B-sqrt}, indeed the coefficient $B$ in the exponent turns out to be exactly the same in both approaches. Moreover, from the linearized solution we can also determine the coefficient $A$ up to order $\mathcal{O}(1/M_s^2)$, i.e. $A=-2v-8v^3\lambda/M_s^2,$ which could not be determined through the asymptotic analysis in Sec.~\ref{subsec-asym} (see Eq.~\eqref{eq:def for asymp sol}).

\begin{figure}[t!]
	\centering
	\includegraphics[scale=0.57]{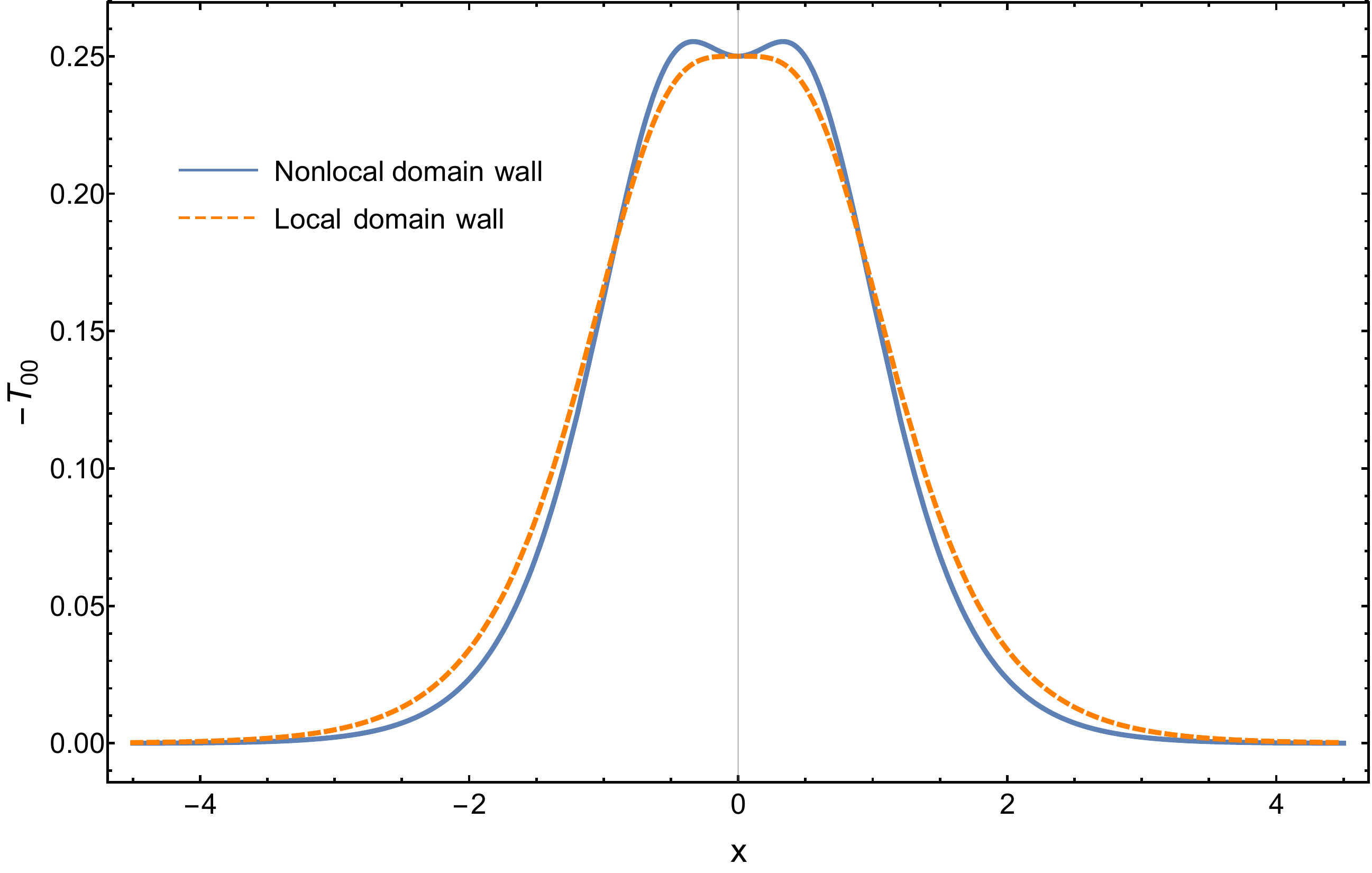}
	\caption{
			In this figure we show the energy density $\mathcal{E}(x)=-T_{00}$ of the linearized nonlocal domain wall solution $\phi(x)=\phi_{\rm L}(x)+\delta\phi(x)$ (solid blue line) in comparison with the one of the local domain wall (orange dashed line). We set $\lambda=2$, $v=1$ and $M_s^2=14.3$.
			}
	\label{fig:energy-density}
\end{figure}

Furthermore, the behavior of the linearized solution close to the origin is quite peculiar as the nonlocal domain wall profile slightly oscillates around the local one. In other words, when going from $x=0$ to $x\rightarrow \infty$ the perturbation $\delta\phi$ is initially negative and then becomes positive; whereas the opposite happens when going from $x=0$ to $x\rightarrow-\infty.$ 
This property may suggest that the typical length scale $\ell$ over which $\phi(x)$ changes in proximity of the origin is larger as compared to the local case. As done for the local domain wall in Sec.~\ref{sec-review}, we can estimate such a length scale as the inverse of the gradient at the origin times the energy scale $v$, i.e.  $\ell\sim v/(\partial_x\phi|_{x=0}).$
By doing so, up to order $\mathcal{O}(1/M_s^2)$ we get
\begin{equation}
\del_x\phi(x)|_{x=0}=v^2\sqrt{\frac{\lambda}{2}}\left(1-\frac{\lambda v^2}{2M_s^2}\right)+\mathcal{O}\left(\frac{1}{M_s^4}\right)\,,\label{grad-zero}
\end{equation}
which yields
\begin{equation}
\ell\sim \sqrt{\frac{2}{\lambda}}\frac{1}{v}\left(1+\frac{\lambda v^2}{2M_s^2}\right)\,.\label{ell-linearized}
\end{equation}
Note that such a length scale does \textit{not} coincide with the width of the wall because it is related to the behavior of the solution close to the origin and far from the vacuum. In standard two-derivative theories the above computation would give a result for $\ell$ that coincides with the size of the wall, but this is just a coincidence. We will comment more on this in Sec.~\ref{subsec-order}.

As a consistency check, we can also compute the energy density of the nonlocal domain wall and confirm that it is positive definite. This must be done consistently by expanding $\mathcal{E}(x)=T^{0}_{0}$ in Eq.~\eqref{eq:energy density for DW} up to order $\mathcal{O}(1/M_s^2).$ In Fig.~\ref{fig:energy-density} we plotted the behavior of the energy density of $\phi(x)$ in comparison with the one of the local kink $\phi_{\rm L}(x).$ As shown in Sec.~\ref{sec-nft-dw}, the positivity of the energy density was a necessary condition to prove that the non-trivial topology ensures the existence of a time-independent non-dissipative solution.

\subsubsection{Validity of the linearized solution}

The above linearized solution was found perturbatively, and it is valid as long as the two inequalities in Eq.~\eqref{pert-loc-nonloc} and~\eqref{ineq-2} are satisfied. We now check when these conditions are verified.

By working with the variable $s=\sqrt{\lambda/2} vx$ and the field redefinition $\delta\phi=vf(s),$ the  inequality~\eqref{pert-loc-nonloc} reads:
\begin{equation}
  |H(s)|= \left|\frac{f(s)}{\tanh s}\right|\ll 1\,,
\end{equation}
where $H(s):=f(s)/\tanh s\,.$ By analyzing the behavior of $|H(s)|$ we can notice that it is always less than unity, thus supporting the validity of the linearized solution in Eq.~\eqref{eq:exact sol for delta phi}; see the left panel in Fig.~\ref{fig:graph for verifivation of perturbative condition}.

Let us now focus on the inequality~\eqref{ineq-2}. By introducing also in this case the variable $s=\sqrt{\lambda/2}vx$ we can write

\begin{eqnarray}
	\frac{\partial_x^2}{M_s^2}\delta\phi
	&=&\frac{\lambda v^2/2}{M_s^2} v f^{\prime\prime}(s) \nonumber\\[2mm]
	&=&\frac{\lambda v^3/2}{M_s^2} \frac{\lambda v^2}{M_s^2} \frac{{\rm d}^2}{{\rm d}s^2}
		\qty{
			\frac{1}{\cosh^2 s}
			\left(
				\frac{3}{4} \log\frac{1+\tanh s}{1-\tanh s} - 2\tanh s
			\right)
		 } \nonumber\\[2mm]
	&=&v \times
		\underbrace{\frac{1}{2}\qty(\frac{\lambda v^2}{M_s^2})^2\frac{{\rm d}^2}{{\rm d}s^2}
			\qty{
				\frac{1}{\cosh^2 s}
				\left(
					\frac{3}{4} \log\frac{1+\tanh s}{1-\tanh s} - 2\tanh s
				\right)
			}
		}_{=:\,g(s)}\,,
\end{eqnarray}
where $g(s):=\partial_x^2\delta\phi(x)/(M_s^2v)$.
In terms of the dimensionless functions $f(s)$ and $g(s)$ the inequality~\eqref{ineq-2} becomes
\begin{align}
	\left|\frac{\partial_x^2}{M_s^2}\delta\phi\right| \ll |\delta\phi|
		\quad\Leftrightarrow\quad
	|g(s)| \ll |f(s)|\,.
\end{align}
Therefore, we have to analyze the function
\begin{align}
	h(s):=\frac{g(s)}{f(s)}
				=\frac{\lambda v^2}{2M_s^2} 
	 					\frac{\dps\frac{{\rm d}^2}{{\rm d}s^2}\qty[\frac{1}{\cosh^2 s}\qty(\frac{3}{4}\log\frac{1+\tanh s}{1-\tanh s}-2\tanh s)]}
			 						{\dps\frac{1}{\cosh^2 s}\qty(\frac{3}{4}\log\frac{1+\tanh s}{1-\tanh s}-2\tanh s)}\,,
\end{align}
and check for which values of $s$ its modulus $|h(s)|$ is less than unity.
In the right panel of Fig.~\ref{fig:graph for verifivation of perturbative condition} we have shown the behavior of $h(s);$ we have only plotted the region $s\geq 0$ as $h(s)$ is an even function in $s$.
We can notice that $h(s)$ diverges at the point $s\sim 1.03402$ where $f(s)$ vanishes; therefore, in the proximity of this point the linearized solution $\delta\phi(x)$ might not be valid.
However, for $|s|\rightarrow \infty$ and $s\rightarrow 0$ the inequality~\eqref{ineq-2} can be satisfied:

\begin{figure}[t]
	\centering
	\includegraphics[scale=0.315]{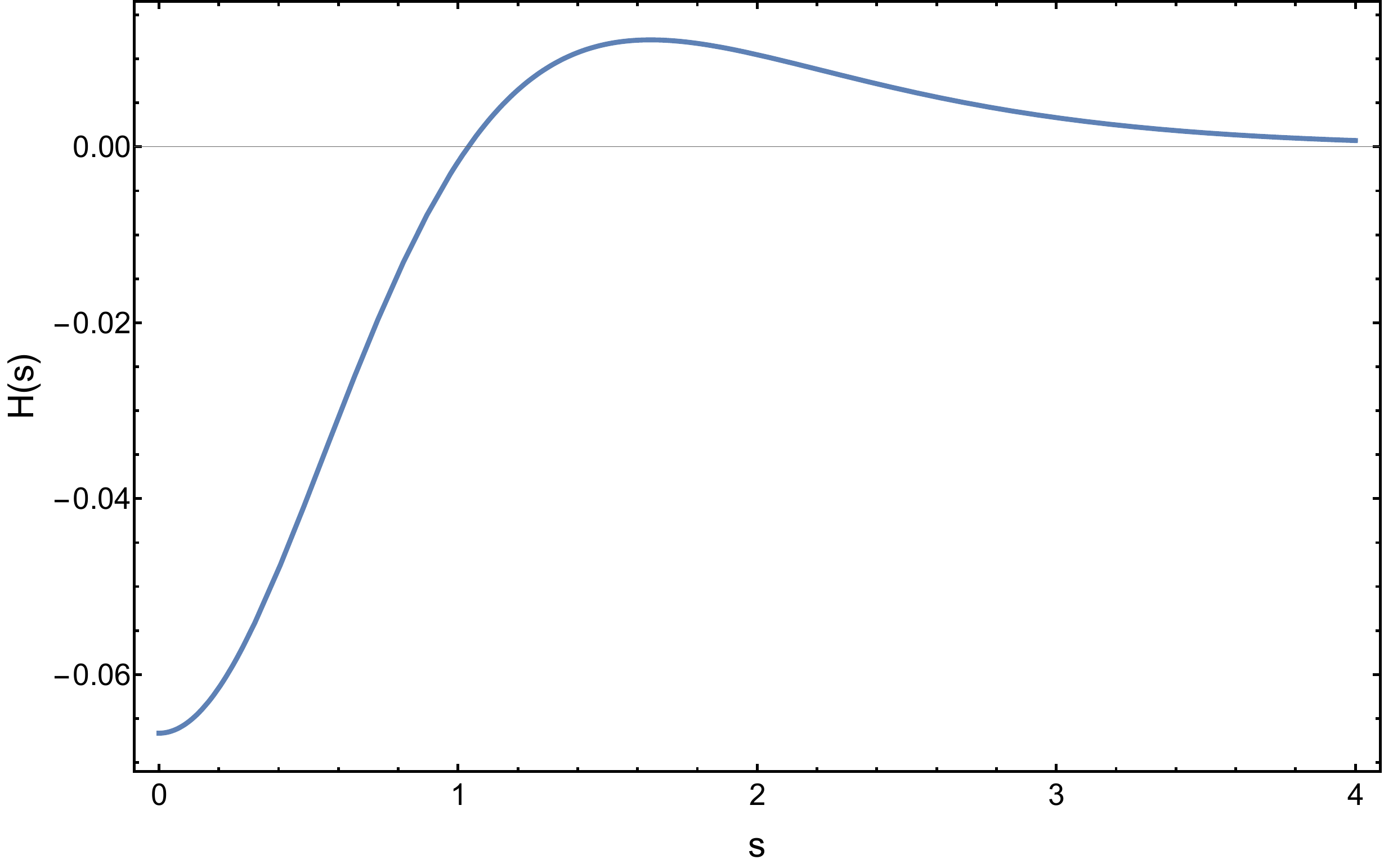}\quad\includegraphics[scale=0.305]{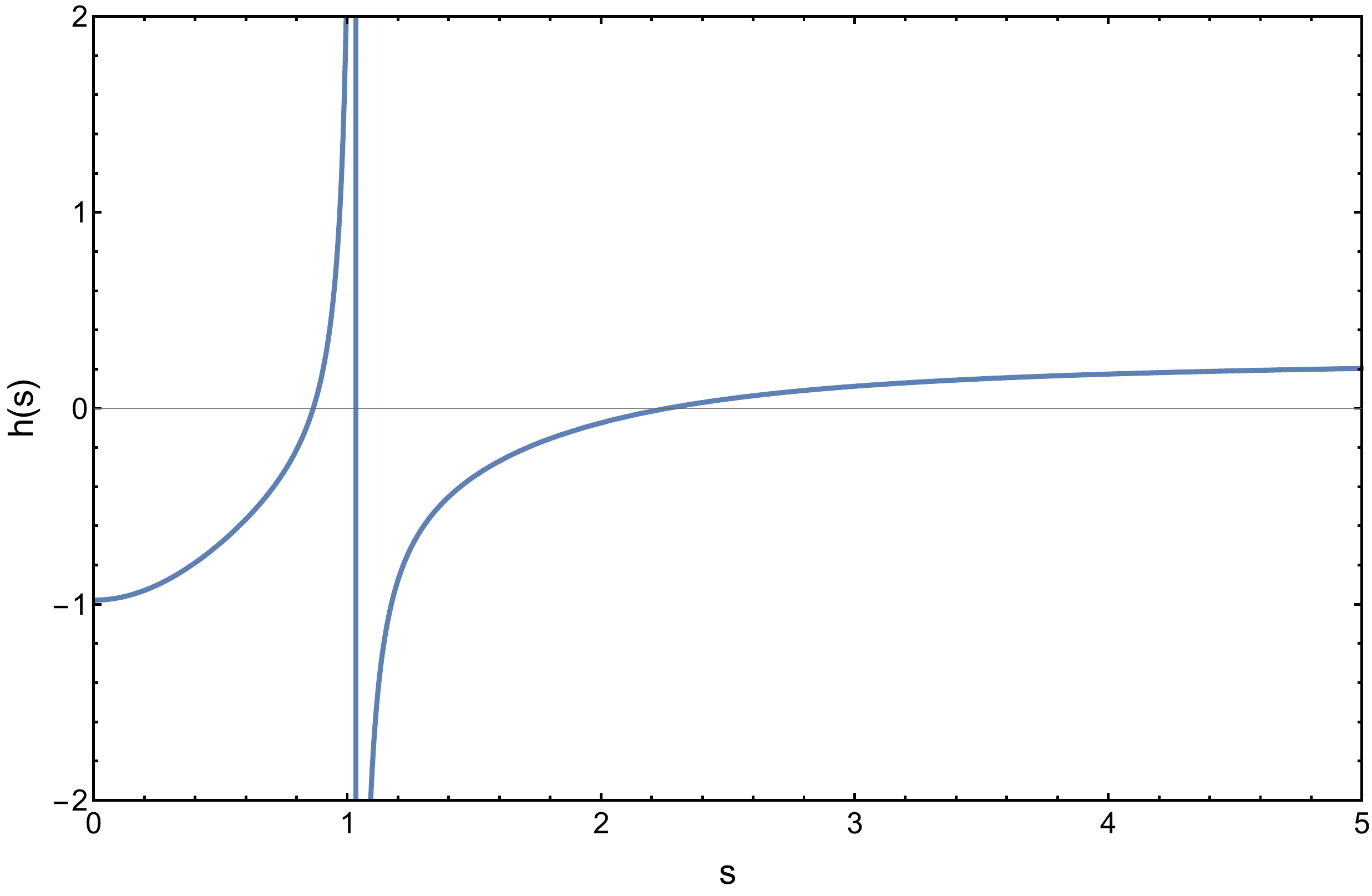}
	\caption{
			(Left panel) behavior of $H(s)=f(s)/\tanh s$ as a function of $s=\sqrt{\lambda/2}vx.$ The modulus of the function is always less than unity, i.e. $|H(s)|< 1,$ supporting the validity of the linearized solution $\delta \phi (x) =v f(x)$ in Eq.~\eqref{eq:exact sol for delta phi}. $H(s)$ becomes smaller and smaller for larger values of $M_s.$   
			(Right panel) behavior of the function $h(s)=g(s)/f(s).$
			As long as $|h(s)|\ll 1$ the linearized solution~\eqref{eq:exact sol for delta phi}  can be trusted as a good approximation of the true behavior of the nonlocal domain wall.
			We can notice that close to the asymptotics ($s\rightarrow \infty$) the function can be kept less than one, but there is a singularity at $s\sim 1.03402$ caused by the fact that $f(s)$ vanishes at this point.
            Moreover, the linearized approximation close to the origin and at infinity becomes better for larger values of $M_s.$
			In both panels we only showed the behavior for $s\geq 0$ because both functions $h(s)$ and $H(s)$ are even in $s.$
			We set $\lambda=2,$ $v=1$ and $M_s^2=14.3,$ which are consistent with the theoretical constraint $M_s^2\geq -\lambda v^2/W_0(-1/3e)$ in Eq.~\eqref{eq:neccessary cond for asymp sol}.
		}
	\label{fig:graph for verifivation of perturbative condition}
\end{figure}

%
\begin{align}
 	h(0)=\lim_{s\to0} h(s) = -7\frac{\lambda v^2}{M_s^2}\,,
 	\qquad
	h(\infty)=\lim_{s\to\infty} h(s) = 2\frac{\lambda v^2}{M_s^2}\,,\label{limits}
\end{align}
and by using the theoretical lower bound in Eq.~\eqref{eq:neccessary cond for asymp sol}, i.e. $\lambda v^2/M_s^2\leq-W_0^{-1}(-1/3e)\sim 0.14,$  it follows that both asymptotic limits are always less than unity, i.e. $|h(0)|< 0.98$ and $h(\infty)< 0.28,$ and the approximation becomes better for larger values of the scale of nonlocality $M_s$.

Let us now make two important remarks.

\paragraph{Remark 3.} In light of the remark at the end of Sec.~\ref{subsec-asym} we now understand that the linearized perturbative solution~\eqref{eq:exact sol for delta phi} would have not been valid if instead we had used $W_{-1}$ and the corresponding upper bound $M_s\leq -\lambda v^2/W_{-1}(-1/3e).$ In such a case a different domain wall solution with the same functional form of the asymptotic behavior is obviously guaranteed to exist, but we are not interested in it in the current work. Therefore, we emphasize again that we only work with a domain wall configuration consistent with the bound~\eqref{eq:neccessary cond for asymp sol}.

\paragraph{Remark 4.} We have noticed that the linearized approximation breaks down in the proximity of $s\sim 1.03402.$ It means that the boundary condition $\delta\phi(\pm\infty)=0$ imposed on the perturbation cannot be used to connect the behavior of the solution from $x=\pm\infty$ to $x=0$ because the boundary condition itself would break down. This might imply that apparently the linearized solution obtained in Eq.~\eqref{eq:exact sol for delta phi} can be trusted only close to the vacua, and that the above analysis is not enough to justify the behavior close to the origin.
However, by using a different and reliable perturbative expansion in the intermediate region around $s\sim 1.03402,$ and imposing junction conditions to glue different pieces of the solution defined in three different regions, we checked and confirmed that the behavior close to the origin found above is well justified. See App.~\ref{sec-corr} for more details.

\subsection{Estimation of width and energy} \label{subsec-order}

We now estimate the width and the energy per unit area of the nonlocal domain wall by performing an analogous analysis as the one made for the local two-derivative case in Sec.~\ref{sec-review}.

Let us approximate the field as $\phi\sim  v,$  and the gradient as $\del_x\sim R^{-1}$ where $R$ is the width of the wall.
The next step would be to impose the balance between the kinetic and the potential energy in Eq.~\eqref{eq:energy density for DW} for the lowest-energy configuration, and solve the resulting equation for the width $R.$ For convenience, we recall the expression for the energy density~\eqref{eq:energy density for DW}
\begin{align}
	\mathcal{E}(x)= -\frac{1}{2}\phi e^{-\partial_x^2/M_s^2}(\partial_x^2+\lambda v^2)\phi
			+\frac{\lambda}{4}(\phi^4+v^4)\,.
	\label{eq:energy density for DW-2}
\end{align}
The presence of the infinite-derivative operator makes the procedure less straightforward as compared to the local case because we should first understand how to estimate $e^{-\del_x^2/M_s^2},$ i.e. whether to replace the exponent with $-1/M_s^2R^2$ or $+1/M_s^2R^2.$ 

The strategy to follow in order to avoid any ambiguities is to Taylor expand, recast the infinite-derivative pieces in terms of an infinite number of squared quantities, replace $\del_x\sim 1/R,$ $\phi\sim v,$ and then re-sum the series.

By Taylor expanding in powers of $\partial_x^2/M_s^2$ the infinite-derivative terms in Eq.~\eqref{eq:energy density for DW-2}, and neglecting total derivatives, we can write
\begin{align}
\!\phi e^{-\del_x^2/M_s^2}\del_x^2\phi&= -\del_x\phi e^{-\del_x^2/M_s^2}\del_x\phi\nonumber\\[2mm]
&= -\left[(\del_x\phi)^2-\frac{1}{M_s^2}\del_x\phi\del_x^2\del_x\phi+\frac{1}{2!M_s^4}\del_x\phi\del_x^4\del_x\phi-\cdots+\frac{(-1)^n}{n!M_s^{2n}}\del_x\phi\del_x^{2n}\del_x\phi+\cdots\right]  \nonumber\\[2mm]
&=-\left[(\del_x\phi)^2+\frac{1}{M_s^2}(\del_x^2\phi)^2+\frac{1}{2!M_s^4}(\del_x^3\phi)^2+\cdots+\frac{1}{n!M_s^{2n}}(\del_x^{n+1}\phi)^2  +\cdots\right]\nonumber\\[2mm]
&= -\sum\limits_{n=0}^\infty \frac{1}{n!}\left(\frac{1}{M_s^2}\right)^n \left(\del_x^{n+1}\phi\right)^2\,,
\end{align}
and
\begin{equation}
\lambda v^2\phi e^{-\del_x^2/M_s^2}\phi=\lambda v^2\sum\limits_{n=0}^\infty \frac{1}{n!}\left(\frac{1}{M_s^2}\right)^n \left(\del_x^{n}\phi\right)^2\,.
\end{equation}
Then, by using $\phi\sim v$ and $\del_x\sim 1/R,$ we get
\begin{align}
\phi e^{-\del_x^2/M_s^2}\del_x^2\phi\sim -\frac{v^2}{R^2}\sum\limits_{n=0}^\infty\frac{1}{n!}\left(\frac{1}{M_s^2R^2}\right)^n = -\frac{v^2}{R^2}e^{1/(M_sR)^2}\,,
\end{align}
and
\begin{equation}
\lambda v^2\phi e^{-\del_x^2/M_s^2}\phi\sim\lambda v^4 e^{1/(M_sR)^2}\,.
\end{equation}
Thus, we have shown that the correct sign in the exponent when making the estimation is the positive one\footnote{To further remove any possible ambiguity and/or confusion, it is worth mentioning that the same result would have been obtained if we would have started with a positive definite expression for the kinetic energy, for instance with the expression $\del_x\phi e^{-\del_x^2/M_s^2}\del_x\phi=(e^{-\del_x^2/2M_s^2}\del_x\phi)^2\geq 0,$ where we integrated by parts and neglected total derivatives. Also in this case one can show (up to total derivatives) that $(e^{-\del_x^2/2M_s^2}\del_x\phi)^2=\sum_{k,l=0}^\infty 1/(k!\,l!) (1/2M_s^2)^{k+l}\left(\del_x^{(k+l+1)}\phi\right)^2\sim (v^2/R^2)\left[\sum_{k=0}^\infty 1/k!(1/2M_s^2R^2)^n\right]^2= (v^2/R^2)e^{1/(M_sR)^2}$.}.

To make more manifest the consistency with the low-energy limit $M_s \to \infty$, it is convenient to separate the kinetic and the potential contributions in~\eqref{eq:energy density for DW-2} as follows:
\begin{align}
\mathcal{E}(x)=&
    \qty[
        -\frac{1}{2} \phi e^{-\del_x^2/M_s^2}\partial_x^2 \phi
        - \frac{1}{2} \lambda v^2 \phi \qty(e^{-\del_x^2/M_s^2}-1) \phi
    ]
    + \qty[
        \frac{\lambda}{4} \qty(\phi^2-v^2)^2
    ]\,,
\end{align}
so that the balance equation between  kinetic and  potential energies reads
\begin{align}
    \frac{1}{2} \frac{v^2}{R^2} e^{1/(M_s R)^2} - \frac{1}{2} \lambda v^4 \qty(e^{1/(M_s R)^2}-1)
        \sim
    \frac{\lambda}{4} v^4\,.
    \label{eq:full-balance-eq}
\end{align}
We are mainly interested in the leading nonlocal correction, thus we expand for $M_sR\gg 1$ up to the first relevant nonlocal contribution:
\begin{eqnarray}
 \frac{1}{2}\frac{v^2}{R^2}\left(1+\frac{1}{M_s^2R^2}\right)-\frac{1}{2}\frac{\lambda v^4}{M_s^2R^2}\sim \frac{1}{4}\lambda v^4\nonumber
	\label{eq:full-balance-eq-leading}
\end{eqnarray}
The solution up to order $\mathcal{O}(1/M_s^2)$ is given by
\begin{eqnarray}
\frac{1}{R^2}\sim\frac{\lambda v^2}{2}\left(1+\frac{\lambda v^2}{2M_s^2} \right)\,,
\end{eqnarray}
from which we obtain
\begin{eqnarray}
	R\sim\sqrt{\frac{2}{\lambda }}\frac{1}{v}\left(1-\frac{\lambda v^2}{4M_s^2} \right)\,.
\end{eqnarray}
Therefore, the width of the nonlocal domain wall, $R$, turns out to be \textit{thinner} as compared to the local two-derivative case. This is consistent with both the asymptotic analysis in Sec.~\ref{subsec-asym} and with the linearized solution in Sec.~\ref{subsec-pert}. In fact, in the previous Subsections we found that the nonlocal configuration approaches the vacua $\pm v$ \textit{faster} as compared to the local case, i.e. the coefficient $B$ in Eq.~\eqref{eq:series expansion for coeff B-sqrt} is larger than the corresponding local one. Then, as shown in Eq.~\eqref{eq:tildeR}, $\widetilde{R}$ becomes smaller in the nonlocal case, which is consistent with the behaviour of $R$. That is, the coefficient $B$ and the width of the wall $R$ should be inversely proportional to each other; this means that if $B$ increases then $R$ must decrease, and indeed this is what we showed.

We can also estimate the energy per unit area 
\begin{eqnarray}
	E&=& \int_\mathbb{R}{{\rm d}x}\qty[-\frac{1}{2}\phi e^{-\partial_x^2/M_s^2}(\partial_x^2+\lambda v^2)\phi
			+\frac{\lambda}{4}(\phi^4+v^4)] \nonumber\\[2mm]
	 &\sim& (\text{width of the wall}) \times (\text{energy density}) \nonumber\\[2mm]
		& \sim& R \times \lambda v^4 \nonumber\\[2mm]
		&\sim& \sqrt{\frac{\lambda}{2}}v^3\left(1-\frac{\lambda v^2}{4M_s^2}\right) \,,\label{energ-estim}
\end{eqnarray}
which is also \textit{decreased} as compared to the local case.

It is worth to emphasize that the expansion for small $\lambda v^2/M_s^2$ is well justified for the domain wall solution satisfying the bound in Eq.~\eqref{eq:neccessary cond for asymp sol} which was obtained by using the principal branch $W_0$ of the Lambert-W function.

\paragraph{Remark 5.} In Sec.~\ref{subsec-pert} we have estimated an additional scale $\ell$ in addition to the width ($\ell> R, \widetilde{R}$). In standard local theories all of the three scales are the same because there is only one physical scale, $\ell_{\rm L}=R_{\rm L}=\widetilde{R}_{\rm L} \sim (\sqrt{\lambda}v)^{-1}.$  In fact, in general $\ell$ and $R$ ($\widetilde{R}$) represent two different physical scales, and this becomes manifest in the nonlocal theory under investigation. The length scale $R$ ($\widetilde{R}$) is the one that contains the information about the size of the wall because it is proportional to $1/B,$ and it is related to how fast the field configuration approaches the vacuum. Whereas, the scale $\ell$ is related to how fast the field changes in the proximity of the origin, indeed it is inversely proportional to the gradient at $x=0$ (see Eqs.~\eqref{grad-zero} and~\eqref{ell-linearized}), and we have $\ell>\ell_{\rm L}.$ The difference between $\ell$ and $R$ ($\widetilde{R}$) is caused by the oscillatory behavior of the nonlocal solution around the local one.


\section{Comments on other topological defects} \label{sec-other}

So far we have only focused on the domain wall configuration. However, it would be very interesting if one could repeat the same analysis also in the case of other topological defects, like \textit{string} and \textit{monopole} which can appear in nonlocal models characterized by continuous-symmetry breaking. A full study of these topological defects in nonlocal field theories goes beyond the scope of this paper, however we can make some important comments.

First of all, the existence of such finite-energy configurations is always guaranteed by the non-trivial topological structure of the vacuum manifold.\footnote{Of course, they can exist only dynamically in the local case because of Derrick's theorem~\cite{Vilenkin:2000jqa}. But, in a nonlocal case, even Derrick's theorem might be circumvented. This issue will be left for a future work.}
Knowing that a solution must exist, then we could ask how some of their properties would be affected by nonlocality. Actually, a similar order-of-magnitude estimation as the one carried out in Sec.~\ref{subsec-order} can be applied to these other global topological defects. In particular, by imposing the balance between kinetic and potential energy one would obtain that the radius of both string and monopole are \textit{smaller} as compared to the corresponding ones in the local case. 

We leave a more detailed investigation of higher dimensional topological defects,  including stabilizing gauge fields, for future tasks.


\section{Discussion \& conclusions}\label{sec-dis}

%
%
\paragraph{Summary.} In this paper, we studied for the first time topological defects in the context of the nonlocal field theories. In particular, we mainly focused on the domain wall configuration associated to the $\mathbb{Z}_2$-symmetry breaking in the simplest nonlocal scalar field theory with nonlocal differential operator $e^{-\Box/M_s^2}$. Despite the complexity of non-linear infinite-order differential equations, we managed to find an approximate analytic solution. Indeed, we were able to understand how nonlocality affects the behavior of the domain wall both asymptotically close to the vacua and around the origin. The consistency of different methods to find the solution confirm that our linearized treatment is mathematically correct.

Let us briefly highlight our main results:
\begin{itemize}
    
    \item We showed that the nonlocal domain wall approaches the asymptotic vacua $\pm v$ faster as compared to the local two-derivative case. We confirmed this feature in two ways: (i) studying the behavior of the solution towards infinity ($|x|\rightarrow \infty$); (ii) analyzing a linearized nonlocal solution found through perturbations around the local domain wall configuration. 
    
    \item Such a faster asymptotic behavior also means that the width of the wall, $R$ ($\widetilde{R}$), is smaller than the corresponding local one. This physically means that the boundary separating two adjacent casually disconnected spatial regions with two different vacua (i.e. $+v$ and $-v$) becomes thinner as compared to the local case. We confirmed this property by making an order-of-magnitude estimation involving the balance equation between kinetic and potential energy. As a consequence, also the energy per unit area can be shown to be smaller.

    \item We noticed that the nonlocal domain wall has a very peculiar behavior around the origin, i.e. in the proximity of $x\sim 0$. We found that the linearized nonlocal solution, $\phi=\phi_{\rm L}+\delta \phi,$ oscillates around the local domain wall when going from $x=0$ to $|x|\rightarrow \infty.$ In other words, the perturbation $\delta\phi$ changes sign: when going from $x=0$ to $x=+\infty$ it is first negative and then positive, and vice-versa when going from $x=0$ to $x=-\infty$ . We confirmed the validity of the solution close to the origin in App.~\ref{sec-corr}.

    \item The specific nonlocal domain wall solution analyzed in this paper can exist only if the nonlocal scale $M_s$ satisfies the lower bound $M_s\gtrsim \sqrt{\lambda} v,$ namely if the energy scale of nonlocality is larger than the symmetry-breaking scale.

\end{itemize}

\paragraph{Discussion \& Outlook.}
Here we have only dealt with nonlocal field theories in flat spacetime without assuming any specific physical scenario. However, it might be interesting to understand how to embed our analysis in a cosmological context where we could expect also gravity to be nonlocal; see Refs.~\cite{Biswas:2005qr,Koshelev:2016xqb,Koshelev:2017tvv,Koshelev:2020foq} and references therein. 

In particular, in Refs.~\cite{Koshelev:2017tvv,Koshelev:2020foq} inflationary cosmology in nonlocal (infinite-derivative) gravity was investigated, and the following experimental bound on the scale of nonlocality was obtained: $M_s\gtrsim H,$ where $H$ is the Hubble constant during inflation, i.e. $H\sim 10^{14}$GeV. 
Our theoretical lower bound is consistent with the experimental constraint derived in~\cite{Koshelev:2017tvv,Koshelev:2020foq} for the gravity sector. Indeed, some symmetry breaking is expected to happen after inflation, i.e. at energies $v\lesssim H,$ which is consistent with the theoretical lower bound $M_s\gtrsim \sqrt{\lambda} v$ in Eq.~\eqref{eq:neccessary cond for asymp sol}.

Hence, in a cosmological context one would expect the following hierarchy of scales\footnote{We are implicitly assuming that there exists only one scale of nonlocality $M_s$ for both gravity and matter sectors.}:
\begin{equation}
M_s\gtrsim H\gtrsim v\,.\label{set-ineq}
\end{equation}
Very interestingly, this cosmological scenario can be used to rule out some topological-defect solutions in nonlocal field theory. For instance, in light of the discussions a the end of Sec.~\ref{subsec-asym} and Sec.~\ref{subsec-pert}, there must exist at least another domain wall configuration that is valid for $M_s^2\lesssim \lambda v^2.$ In such a case the set of inequalities~\eqref{set-ineq} would be replaced by $v\gtrsim M_s\gtrsim H,$ which implies that the symmetry breaking would happen before inflation. Thus, if we are interested in domain wall formation after inflation, then we can surely discard any configuration valid in the regime $M_s^2\lesssim \lambda v^2.$ 

It would be very interesting if one would consider other topological defects, like strings and monopoles, which can appear in nonlocal models characterized by continuous-symmetry breaking. In fact, global string might play important roles like axion emissions in an expanding universe. In the local case, the topological defects that are formed from global-symmetry breaking should be unstable because of Derrick's theorem~\cite{Vilenkin:2000jqa} which excludes the existence of stationary stable configurations in dimensions greater than one. Then, they can exist only dynamically e.g. in an expanding universe. However, Derrick's theorem might not apply to a nonlocal case thanks to the nonlocality. It would be also interesting to investigate whether such stationary stable configurations could exist in a nonlocal case or not.

As another potential future direction to follow we can consider another class of models characterized by gauge symmetries as well as global ones. In fact, among the possible physical applications that can be studied in relation to topological defects, we have gravitational waves, e.g. the ones emitted by cosmic strings. We would expect that the presence of nonlocality would change the dynamics in such a way to modify non-trivially the gravitational wave-form.
This type of investigations will provide powerful test-benches to test nonlocal field theories, and to further constrain the structure of the nonlocal differential operators in the Lagrangian and the value of the nonlocal scale. In this work we only focused on the simplest model with $F(-\Box)=e^{-\Box/M_s^2},$ but one could work with more generic operators. Actually, the class of viable differential operators is huge (e.g. see~\cite{Buoninfante:2020ctr}), and it would be interesting to reduce it by means of new phenomenological studies.

More generally, one can consider gravitational effects sourced by nonlocal topological defects in both local and nonlocal theories of gravity; see Refs.~\cite{Boos:2018bxf,Kolar:2020bpo,Boos:2020kgj,Boos:2021suz,Buoninfante:2021wuw} for studies concerning nonlocal gravitational fields sourced by local topological defects and other local extended objects.

As yet another future work we also wish to consider other type of field theoretical objects. An important phenomenon is the \textit{false vacuum decay}~\cite{Coleman:1977py,Callan:1977pt,Coleman:1980aw} according to which the false vacuum -- which corresponds to a local minimum of the potential -- has a non-zero probability to decay through quantum tunneling into the true vacuum -- which corresponds to a global minimum. This tunneling process consists in interpolating between the false and the true vacuum through an instanton (a bounce solution).
It may be very interesting to generalize the standard analysis done for a local two-derivative theory to the context of nonlocal field theories, and to understand how nonlocality would affect this phenomenon, e.g. how the tunneling probability would change. Another interesting direction is to consider non-topological solitons like Q-balls and oscillons/I-balls. The existence of these objects is also related to the presence of a bounce solution. But, one should notice that, different from topological defects, the presence of these bounce solutions is not guaranteed in nonlocal theories. In fact, it is very difficult to guarantee the presence of such a bounce solution in a nonlocal theory, different from the local cases. Therefore, even if one would obtain (possible) approximate solutions somehow, one cannot make any argument based on such approximate solutions without the proof of the existence of exact bounce solutions. This is the reason why we dealt only with topological defects in this paper, and left a study of non-topological field theoretical objects for future work.

Finally, we should emphasize that non-linear and infinite-order differential equations are not only difficult to solve analytically but even numerically. Indeed, up to our knowledge no numerical technique to find domain wall solutions is currently known. Some techniques to solve nonlinear equations involving infinite-order derivatives have been developed in the last decades~\cite{Moeller:2002vx,Arefeva:2003mur,Volovich2003,Joukovskaya:2008cr,Calcagni:2008nm,Frasca:2020ojd}, but none of them seem to be useful for the type of field equations considered in this paper. Previous numerical studies only focused on time-dependent systems for which the differential operator is $e^{+\partial_t^2},$ whereas we are interested in time-independent configurations with $e^{-\partial_x^2}.$ The difference in sign in the exponent is crucial and does not allow us to use the convolution techniques implemented in Refs.~\cite{Moeller:2002vx,Arefeva:2003mur,Volovich2003,Joukovskaya:2008cr}. Therefore, as a future task it will be extremely interesting to develop new numerical and analytic methods to find topological-defect solutions. This will also be important to investigate the stability of topological defects in nonlocal field theory, something that we have not done in this paper.


\subsection*{Acknowledgements}

The authors are grateful to Sravan Kumar for discussions, and to Carlos Heredia Pimienta and Josep Llosa for useful comments.
Y.~M. acknowledges the financial support from the Advanced Research Center for Quantum Physics and Nanoscience, Tokyo Institute of Technology, and JSPS KAKENHI Grant Number JP22J21295.
M.~Y. acknowledges financial support from JSPS Grant-in-Aid for Scientific Research No. JP18K18764, JP21H01080, JP21H00069. Nordita is supported in part by NordForsk.


\appendix


\section{Behavior of the solution close to the origin} 
\label{sec-corr}

At the end of Sec.~\ref{subsec-pert} we noticed that the behavior of the linearized solution close to the origin was \textit{not} well justified because the linear approximation fails in an intermediate region around $s\sim 1.03402.$
In App.~\ref{subsec-junc} we establish a formalism to justify the linearized solution even in the proximity of the origin; in App.~\ref{subsec-origin} we make a further consistency check by implementing a series expansion method.


\subsection{Nonlocal corrections and junction conditions} \label{subsec-junc}

The idea is to replace the expansion $\phi=\phi_{\rm L}+\delta\phi$ with a different one around $s\sim 1.03402,$ in such a way that a linear approximation can be valid also in this intermediate region. 
We will study the solution in three different regions and in each of them we will find a solution depending on some integration constants to be fixed through boundary and junction conditions. To glue the three solutions we need to choose two points on the left and on the right of $s\sim 1.03402,$ respectively, and show that the resulting full solution is independent of the chosen points as long as the linear approximation remains valid in all three regions. In what follows we perform the analysis by choosing the points $s=0.6$ and $s=1.4,$ but we will also comment on different choices. It will be useful to work with the dimensionless variable $s \equiv \sqrt{\lambda/2}vx$ and function $f(s) \equiv \delta\phi/v$.

\begin{description}
\item[{\bf 1. Close to the vacuum ($s\geq1.4$): }]~\\
    In this region, we use the perturbation $\phi=\phi_{\rm L}+\delta\phi.$ We know that the linearized solution is the one obtained in Sec.~\ref{subsec-pert} by imposing the boundary condition $\delta\phi(\infty)=vf(\infty)=0$ which fixes the integration constant equal to $C_2=-\lambda v^2/16M_s^2.$ Let us call this solution
    \begin{align}
        f_1(s)=\frac{\lambda v^2}{M_s^2}\frac{1}{\cosh^2 s}
			\qty[\frac{3}{4}\log\frac{1+\tanh s}{1-\tanh s} - 2\tanh s]\,.
    \label{sol-reg-1}
    \end{align}
    From Sec.~\ref{subsec-pert} we know that it respects the linear approximation for any $s>1.4$.
    
\item[{\bf 2. Intermediate region ($0.6 \leq s \leq 1.4$): }]~\\
    In this region the solution $f_1(s)$ is not valid, therefore we will use a different linear expansion. We consider 
    \begin{align}
    \delta\phi = \phi - v\,,
    \end{align}
    so that the corresponding linearized equation reads 
    \begin{align}
    e^{-\del_x^2/M_s^2} (\del_x^2+\lambda v^2)\delta\phi = 3 \lambda v^2 \delta\phi\,.
    \end{align}
    Because of the exponential differential operator this equation is very difficult to solve. Alternatively, we can expand the exponential up to some derivative order, and thus solve a higher-order derivative equation:

\begin{align}
e^{-\del_x^2/M_s^2} (\del_x^2+\lambda v^2)\delta\phi
&=\qty(1-\frac{\del_x^2}{M_s^2}+\frac{1}{2}\frac{\del_x^4}{M_s^4}-\frac{1}{3!}\frac{\del_x^6}{M_s^6}+\cdots) (\del_x^2+\lambda v^2)\delta\phi= 3\lambda v^2 \delta\phi\,,
\label{expand-exp-for-inter}
\end{align}
where the dots stand for higher-order derivative terms; we call its solution $f_2(s)=\delta \phi/v.$ Note that the higher the derivative order is, the larger the number of integration constants will be. This implies that a larger number of junction conditions will have to be imposed to determine the full solution. This also means that the accuracy of the final solution will be higher. In particular, we proceed as follows: if we truncate to some derivative order $2n,$ we impose $2n$ junction conditions at the point $s=1.4$ to glue region 1. and region 2., namely we impose the continuity $f_2(1.4)=f_1(1.4)$ and $2n-1$ junctions for the derivatives of order $1,2,\dots,2n-1.$

\item[{\bf 3. Close to the origin ($s \leq 0.6$): }]~\\
     In this region the perturbation $\phi=\phi_{\rm L}+\delta\phi$ is still a valid one, but we cannot use the boundary condition $\delta\phi(\infty)=0$ to fix the integration constant. As explained in Sec.~\ref{subsec-pert}, the reason for this is that the boundary condition breaks down because of the failure of the approximation around $s\sim 1.03402,$ so that we cannot connect the solution from $s=+\infty$ to $s=0.$ 
     
     However, in this region we can still consider the solution
    %
    \begin{eqnarray}
   \!\!\!\!\!\!\!\!\!f_3(s)\!\!\!&=&\!\!\!
     \frac{3C_3/2+27\lambda v^2/32M_s^2}{\cosh^2s} \log\frac{1+\tanh s}{1-\tanh s}
    \nonumber\\[2mm]
    	&&\!\!\!+\left[
    		\qty(2C_3+\frac{\lambda v^2}{8M_s^2})\cosh^2s
    			+ \qty(3C_3-\frac{61\lambda v^2}{16M_s^2})
    			+ \frac{2\lambda v^2}{M_s^2}(1+\tanh^2 s)
    	\right] \tanh s\,,\,\,\,\,\,\,\,\,
    \label{sol-reg-3}
    \end{eqnarray}
    where the integration constant $C_3$ must be determined by imposing the junction condition with the solution in the intermediate region at the point $s=0.6$. In particular, we have to impose the continuity $f_3(0.6)=f_2(0.6).$
\end{description}
\begin{figure}[t!]
	\centering
	\includegraphics[scale=0.372]{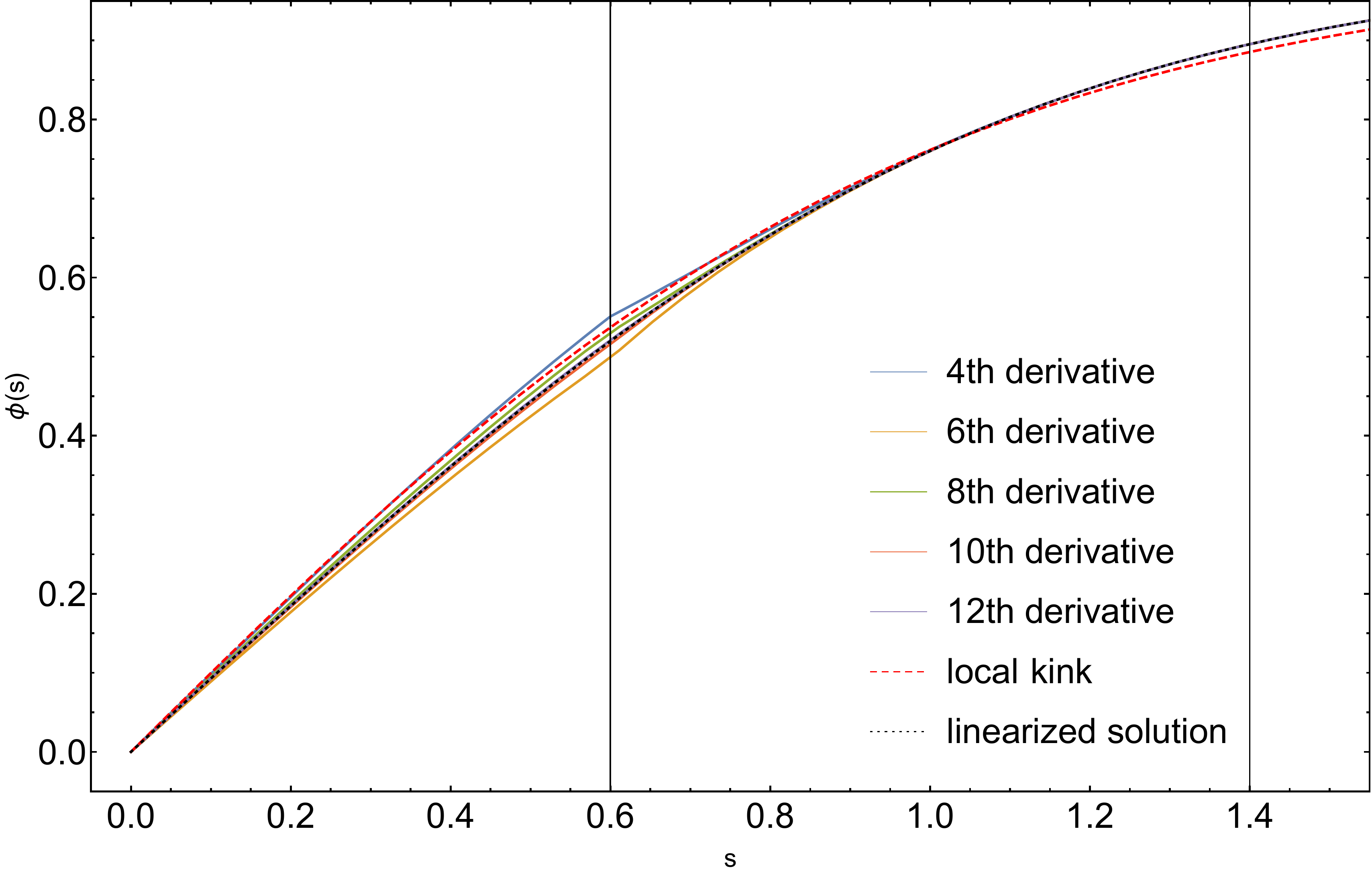}\\[5.5mm]
	\includegraphics[scale=0.37]{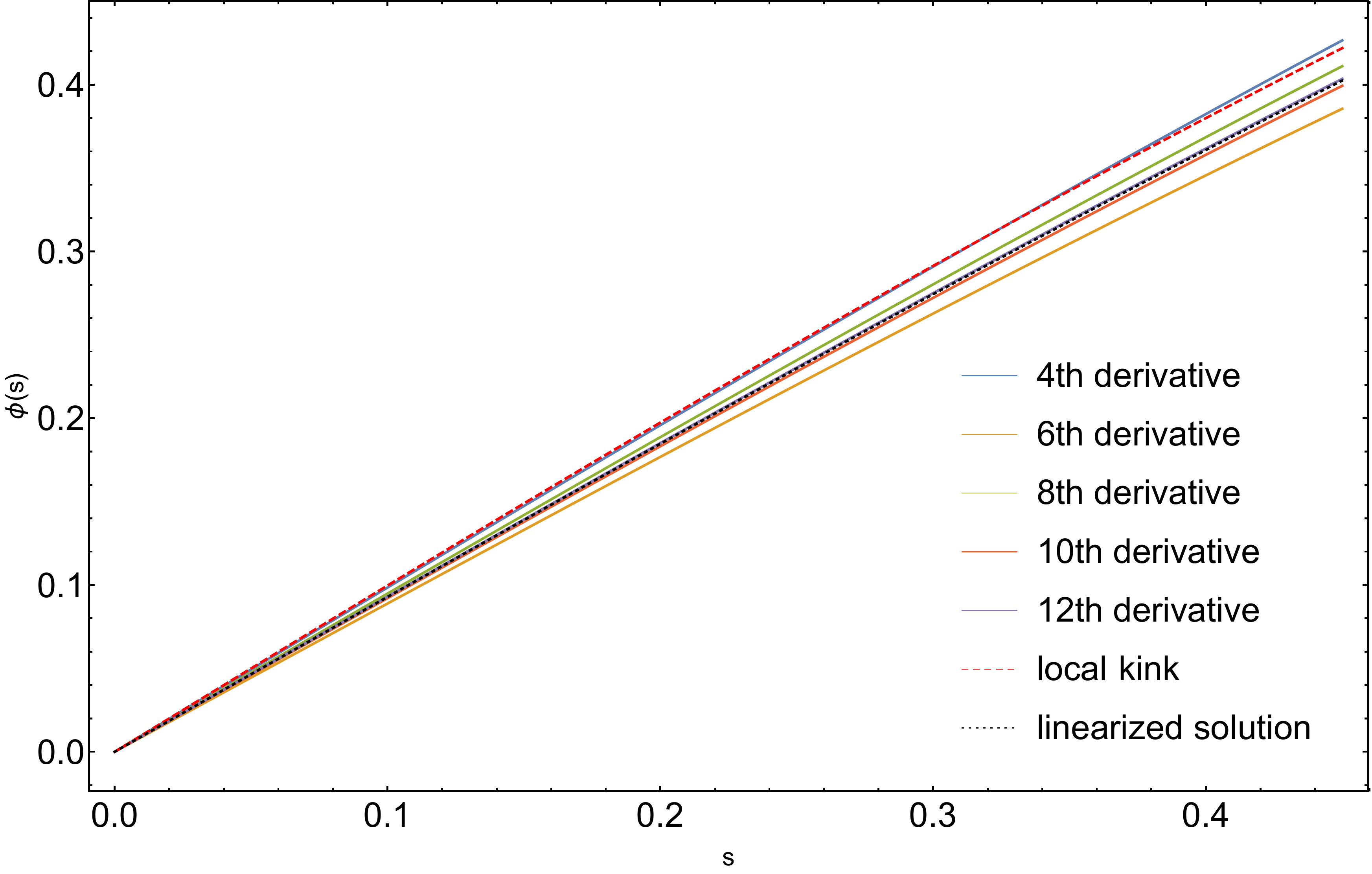}
	\caption{(Top panel) We have reported the behavior of the full glued solution for several higher-derivative truncation of the differential equation in region 2. We performed computations up to $22$nd order derivative, but it is enough to show the results up to $12$th order. The solutions were obtained by imposing the junction conditions at $s=0.6$ and $s=1.4$ (vertical lines) following the procedure explained in this Appendix. 
	The $4$, $6$, $8$, $10$, and $12$-th order cases are plotted in comparison with the linearized solution $\phi=\phi_{\rm L}+\delta\phi$ obtained in Eq.~\eqref{eq:exact sol for delta phi} and the local kink solution. The color and the style for each curve is summarized in the legend. (Bottom panel) we have zoomed on a shorter interval of the $x$-axis in proximity of the origin in order to see more clearly the differences between the curves. 
	In both panels we set $\lambda=2,$ $v=1$ and $M_s^2=14.3,$ which are consistent with the theoretical constraint $M_s^2\geq -\lambda v^2/W_0(-1/3e)$ in Eq.~\eqref{eq:neccessary cond for asymp sol}.
		}
	\label{fig:appended solution with higher derivative case}
\end{figure}
%


We have implemented the above procedure by expanding the solution in the intermediate region up to $22$nd order derivative, and found the corresponding solution $f_2(s)$ numerically. In Fig.\ref{fig:appended solution with higher derivative case}, we showed the behavior of the full glued solution (in all three regions) for $4$, $6$, $8$, $10$ and $12$-th derivative cases, by setting $\lambda=2,\,v=1$ and $M_s^2=14.3$.
One can notice that by increasing the derivative order for the intermediate solution, the behavior close to the origin converges very quickly to the behavior of the linearized solution found in Sec.~\ref{subsec-pert}. This means that the value $B_3=-\lambda v^2/16M_s^2$ is very well justified, and that the linearized solution found in~\eqref{eq:exact sol for delta phi} is a good approximation for the nonlocal domain wall configuration in all three regions. We have also checked that the linear approximation in the region 2. is respected, i.e. $|f_2(s)|\ll1.$
This analysis also confirms that the nonlocal solution is smaller than the local one close to the origin, which physically means that the width of the wall is larger in the nonlocal case.

One might wonder whether the same result holds by choosing other matching points different from $s=0.6$ and $s=1.4$. In fact, we checked that our result is independent of the specific chosen points. For instance, for the pairs of points $(s=0.7,1.5)$ and $(s=0.8,1.6)$ we obtained the same result and confirmed the validity of our analysis.


\subsection{Expansion of the solution around the origin} \label{subsec-origin}

We now introduce a general formalism involving Taylor series expansions around the origin to make a further consistency check of the linearized solution obtained in Sec.~\ref{subsec-pert}

Analogously to the local case and because of the topological structure of the vacuum manifold we expect that the nonlocal domain wall solution is also given by an odd function which is regular at the origin. Thus, we can Taylor expand the solution $\phi(x)$ as
\begin{align}
	\phi(x)
		=& \sum_{n=0}^{\infty} a_{2n+1}(M_s x)^{2n+1}\,,
				\label{eq:asymp-phi.for.TD.near.origin}	
\end{align}
where $M_s x$ is dimensionless and $a_{2n+1}$ are coefficients depending on the scale of nonlocality; we can determine their dependency on $M_s$ as follows. By writing $b_{2n+1}=a_{2n+1}M_s^{2n+1},$ we understand that to consistently recover the local case limit we should have $b_{2n+1}=b_{{\rm L},2n+1}(1+\mathcal{O}(1/M_s^2)),$ where $b_{{\rm L},2n+1}$ is the coefficient in the local case and does not depend on $M_s$. This implies that 
\begin{equation}
a_{2n+1}\sim \mathcal{O}\left(\frac{1}{M_s^{2n+1}}\right)\qquad  \forall n\in \mathbb{N}\,.
\end{equation}
These coefficients can in principle be determined through the field  equation~\eqref{eq:the equation of the motion for DW}.
Let us evaluate both left-hand-side (l.h.s.) and right-hand-side (r.h.s.) of Eq.~\eqref{eq:the equation of the motion for DW}.

Since we are working in a static configuration and in one spatial dimension, we can evaluate the l.h.s. by using the very useful property of the Weierstrass transformation:
\begin{align}
	e^{-\del_x^2} \qty[x^n] = H_n\qty(\frac{x}{2}) \,, \label{eq:propety of Weierstrass transformation}
\end{align}
where $H_n(z)$ is the $n$-th Hermite polynomial, we can write
\begin{eqnarray}
	{\rm l.h.s.}&=&e^{-\del_x^2/M_s^2} \qty(\del_x^2+\lambda v^2) \phi\nonumber\\[2mm]
	&=&e^{-\del_x^2/M_s^2} \left\{ \qty(\del_x^2+\lambda v^2)
			\qty[\sum_{n=0}^{\infty} a_{2n+1}(M_s x)^{2n+1}] \right\} \nonumber\\[2mm]
	&=&M_s^2 e^{-\del_x^2/M_s^2}
			\left\{ \qty(\frac{\del^2}{\del(M_s x)^2}+\frac{\lambda v^2}{M_s^2})
			\qty[\sum_{n=0}^{\infty} a_{2n+1}(M_s x)^{2n+1}] \right\} \nonumber\\[2mm]
	&=&M_s^2 \sum_{n=1}^{\infty} (2n+1) (2n) a_{2n+1} H_{2n-1}\qty(\frac{M_s x}{2})
			+ \frac{\lambda v^2}{M_s^2} \sum_{n=0}^{\infty} a_{2n+1} H_{2n+1}\qty(\frac{M_s x}{2}) \nonumber\\[2mm]
	&=&M_s^2 \sum_{n=0}^{\infty} \qty[ (2n+3) (2n+2) a_{2n+3} + \frac{\lambda v^2}{M_s^2} a_{2n+1} ] H_{2n+1}\qty(\frac{M_s x}{2})\,;
\end{eqnarray}
in the last step we have redefined the summation index in the first term as $n \to n + 1$.
Then, by using the series representation of the Hermite polynomials with odd indexes,
\begin{align}
	H_{2n+1}\qty(z) = (2n+1)! \sum_{k=0}^{\infty} \frac{(-1)^{n-k}}{(2k+1)!(n-k)!}(2z)^{2k+1}\,,
\end{align}
we obtain
\begin{align}
{\rm l.h.s.}=M_s^2 \sum_{n=0}^{\infty} \sum_{k=0}^{\infty} \qty[ (2n+3) (2n+2) a_{2n+3} + \frac{\lambda v^2}{M_s^2} a_{2n+1} ]
	\frac{(2n+1)!(-1)^{n-k}}{(2k+1)!(n-k)!} \qty(M_s x)^{2k+1}\,.
	\label{eq:LHS of the equation of the motion in expandion}
\end{align}

Let us now focus on the r.h.s of Eq.~\eqref{eq:the equation of the motion for DW}. We can use the Cauchy product formula,
\begin{align}
	\qty(\sum_{i=0}^{\infty} a_i x^i) \qty(\sum_{j=0}^{\infty} b_j x^j)
		= \sum_{k=0}^{\infty} c_k x^k
	\qquad \text{with} \quad c_k = \sum_{l=0}^{k} a_l b_{k-l}\,,
\end{align}
to write the cubic term as
\begin{eqnarray}
	{\rm r.h.s.}=\lambda\phi^3=\lambda \sum_{n=0}^{\infty}  \sum_{k=0}^{n} \sum_{i=0}^{k}
			a_{2i+1} a_{2(k-i)+1} a_{2(n-k)+1} \qty(M_s x)^{2n+3}\,.
\end{eqnarray}
Hence, the field equation (l.h.s.$=$r.h.s) can be recast in the following form:
\begin{align}
	&\sum_{n=0}^{\infty} \sum_{k=0}^{\infty}
			\left\{
					\qty[ (2n+3) (2n+2) a_{2n+3} + \frac{\lambda v^2}{M_s^2} a_{2n+1} ]
				\frac{(2n+1)!(-1)^{n-k}}{(2k+1)!(n-k)!}
			\right\}\qty(M_s x)^{2k+1}
\nonumber\\[2mm]
&\qquad
			= \sum_{n=0}^{\infty} 
				\left\{
					\lambda \sum_{k=0}^{n} \sum_{i=0}^{k} a_{2i+1} a_{2(k-i)+1} a_{2(n-k)+1}
				\right\}
				\qty(M_s x)^{2n+3}
				\label{field-eq-coeff}
\end{align}
Moreover, by introducing the new index $l$ through $n=k+l,$ we can write the l.h.s. as
\begin{align}
	{\rm l.h.s.}=\sum_{k=0}^{\infty} \sum_{l=0}^{\infty}
	&\left\{
		\left[(2(k+l)+3)(2(k+l)+2) a_{2(k+l)+3}+\frac{\lambda v^2}{M_s^2} a_{2(k+l)+1}\right]
\right.\nonumber\\&\qquad\qquad\qquad\qquad\qquad\qquad\qquad\qquad\left.\times
		\frac{(2(k+l)+1)!(-1)^{l}}{(2k+1)!l!}
	\right\}
	\qty(M_sx)^{2k+1}\,.
\end{align}
By replacing the index $k$ with $n$ and extracting the term proportional to $(M_s x)$, we obtain
\begin{align}
&{\rm l.h.s.} =\qty[
	\sum_{l=0}^{\infty}
	\qty((2l+3)(2l+2) a_{2l+3}+\frac{\lambda v^2}{M_s^2} a_{2l+1})
	\frac{(2l+1)!(-1)^{l}}{l!}
](M_s x)
\nonumber\\[2mm]
&\qquad
+\sum_{n=0}^{\infty} \left[
	\sum_{l=0}^{\infty}
	\left\{
		\qty((2(n+l)+5)(2(n+l)+4) a_{2(n+l)+5}+\frac{\lambda v^2}{M_s^2} a_{2(n+l)+3})
\right.\right.\nonumber\\[2mm]
&\qquad\qquad\qquad\qquad\qquad\left.\left.\times
	\frac{(2(n+l)+3)!(-1)^{l}}{(2n+3)!l!}\right\}
\right]\qty(M_sx)^{2n+3}\,.
\end{align}
We can now factorize the pieces $(M_s x)^{2n+1}$  at any order in $n$ in Eq.~\eqref{field-eq-coeff}, and finally obtain the following algebraic equations for the coefficients:
\begin{align}
\!\!\!&\sum_{l=0}^{\infty}
	\qty((2l+3)(2l+2) a_{2l+3}+\frac{\lambda v^2}{M_s^2} a_{2l+1})
	\frac{(2l+1)!(-1)^{l}}{l!}=0
		\qquad(\text{coefficient of }M_s x)\,,
		\label{eq:coefficient of M_s x in expansion around origin} \\[2.5mm]
\!\!\!& \sum_{l=0}^{\infty}
	\qty((2(n+l)+5)(2(n+l)+4) a_{2(n+l)+5}+\frac{\lambda v^2}{M_s^2} a_{2(n+l+1)+1}) \frac{(2(n+l)+3)!(-1)^{l}}{(2n+3)!l!}  \nonumber\\[1.5mm]
\!\!\!&\qquad\qquad=
	\sum_{k=0}^n
	\qty{\frac{\lambda}{M_s^2}\sum_{i=0}^k a_{2i+1}a_{2(k-i)+1}a_{2(n-k)+1}}
		\qquad(\text{coefficient of } \qty(M_s x)^{2n+1} \text{ with } n\geq1)\,.
		\label{eq:coefficient of higher terms in expansion around origin}
\end{align}
If we could solve the above equations for the coefficients $a_{2n+1},$ we would be able to find an exact solution for $\phi(x).$  

%


\subsubsection{Linearized solution close to the origin}\label{lin-sol-origin}

The formalism introduced above does not really offer a simpler way to find an exact solution.
However, it provides exact relations among the coefficients $a_{2n+1},$ which must be satisfied by any approximate solution that aims at describing the behavior of the domain wall close to the origin.
In other words, the above relations can be used to check the validity of approximate solutions in the region around origin.
In particular, we are going to check whether the coefficients $a_{2n+1}$ of the linearized solution found in Sec.~\ref{subsec-pert} satisfy the relations in Eqs.~\eqref{eq:coefficient of M_s x in expansion around origin} and~\eqref{eq:coefficient of higher terms in expansion around origin} up to $\mathcal{O}(1/M_s^2),$ which is the order in powers of $M_s$ up to which the linearized solution~\eqref{eq:exact sol for delta phi} was determined.  

We Taylor expand $\phi=\phi_{\rm L}+\delta\phi$ (see Eq.~\eqref{eq:exact sol for delta phi})
\begin{equation}
\phi(x)=a_1(M_sx)+a_3(M_sx)^3+a_5(M_sx)^5+a_7(M_sx)^7+a_9(M_sx)^9+\cdots\,,
\end{equation}
with
\begin{eqnarray}
&&a_1=\frac{v^2}{M_s}\sqrt{\frac{\lambda}{2}}\left(1-\frac{\lambda v^2}{2M^2_s}\right)\,,\,\,\,\,\,\,a_3=-\frac{v^4\lambda^{3/2}}{6\sqrt{2}M_s^3}\left(1-\frac{7\lambda v^2}{2M^2_s}\right)\,,\nonumber\\[2mm]
&&a_5=\frac{v^6\lambda^{5/2}}{30\sqrt{2}M_s^5}\left(1-\frac{19\lambda v^2}{2M^2_s}\right)\,,\,\,\,\,\,\,a_7=-\frac{17v^6\lambda^{7/2}}{2510\sqrt{2}M_s^5}\left(1-\frac{635\lambda v^2}{34M^2_s}\right)\,,\nonumber\\[2mm]
&&\qquad\qquad\qquad a_9=\frac{31v^6\lambda^{9/2}}{22680\sqrt{2}M_s^5}\left(1-\frac{1927\lambda v^2}{62M^2_s}\right)\,,
\label{coefficients}
\end{eqnarray}
and substitute the above coefficients into the relations~\eqref{eq:coefficient of M_s x in expansion around origin}  and~\eqref{eq:coefficient of higher terms in expansion around origin}, and verify that they are consistently satisfied up to order $\mathcal{O}(1/M_s^2)$.

The relevant terms up to order $\mathcal{O}(1/M_s^2)$ in Eq.~\eqref{eq:coefficient of M_s x in expansion around origin} are given by
\begin{eqnarray}
M_s^3\left(6a_3+\frac{\lambda v^2}{M_s^2}a_1\right)-6M_s^3\left(20a_5+\frac{\lambda v^2}{M_s^2}a_3\right)+\mathcal{O}\left(\frac{1}{M_s^4}\right)=0\,,
\end{eqnarray}
where we multiplied by $M_s^3$ so that the first term contains contributions of order $\mathcal{O}(1/M_s^0)$ and $\mathcal{O}(1/M_s^2),$ whereas the second term contains the orders $\mathcal{O}(1/M_s^2)$ and $\mathcal{O}(1/M_s^4).$ We can explicitly verify that the orders $\mathcal{O}(1/M_s^0)$ and $\mathcal{O}(1/M_s^2)$ consistently vanish.

At order $\mathcal{O}(1/M_s^0)$ we have
\begin{eqnarray}
M_s^3\left(6a_3+\frac{\lambda v^2}{M_s^2}a_1\right)&=&6\left(-\frac{v^4\lambda^{3/2}}{6\sqrt{2}}\right)+\lambda v^2\left( \frac{v^2\sqrt{\lambda}}{\sqrt{2}}\right)+\mathcal{O}\left(\frac{1}{M_s^2}\right)\nonumber\\[2mm]
&=&0+\mathcal{O}\left(\frac{1}{M_s^2}\right)\,;
\end{eqnarray}
while at order $\mathcal{O}(1/M_s^2)$  we get
\begin{eqnarray}
M_s^3\left(6a_3+\frac{\lambda v^2}{M_s^2}a_1\right)-6M_s^3\left(20a_5+\frac{\lambda v^2}{M_s^2}a_3\right)&=&\frac{7v^6\lambda^{5/2}}{2\sqrt{2}M_s^2}-\frac{v^6\lambda^{5/2}}{2\sqrt{2}M_s^2}\nonumber\\[2mm]
&&-\frac{4v^6\lambda^{5/2}}{\sqrt{2}M_s^2}+\frac{v^6\lambda^{5/2}}{\sqrt{2}M_s^2} + \mathcal{O}\left(\frac{1}{M_s^4}\right)  \nonumber\\[2mm]
&=&0+\mathcal{O}\left(\frac{1}{M_s^4}\right)\,.
\end{eqnarray}
One can verify that also the relation~\eqref{eq:coefficient of higher terms in expansion around origin} is satisfied. For instance, for $n=1$ the relevant terms up to order  $\mathcal{O}(1/M_s^2)$ are
\begin{eqnarray}
M^7\left[\left(42 a_7+\frac{\lambda v^2}{M_s^2}a_5\right)-42\left(72 a_9+\frac{\lambda v^2}{M_s^2}a_7\right)\right]=M^7\left(3\frac{\lambda}{M_s^2}a_1^2a_3\right)\,,\label{eq-n=1}
\end{eqnarray}
where we have multiplied by $M^7$ both sides of Eq.~\eqref{eq:coefficient of higher terms in expansion around origin} to isolate the orders $\mathcal{O}(1/M_s^0)$ and $\mathcal{O}(1/M_s^2);$ now the coefficients $a_7$ and $a_9$ contribute to the analysis. By substituting the expressions~\eqref{coefficients} for the coefficients, we can easily show that the relation~\eqref{eq-n=1} is consistently satisfied up to order $\mathcal{O}(1/M_s^2).$

Hence, by making use of the series-expansion formalism we obtained an additional consistency check for the validity of the linearized domain wall solution found in Sec.~\ref{subsec-pert}.


\section{Energy-momentum tensor in nonlocal field theories} \label{app-em}

In this Appendix we are going to derive a very general expression for the energy momentum tensor in nonlocal field theories by taking into account the presence of infinitely many derivatives in the Lagrangian.

Let us consider the following action 
\begin{align}
	S = \int d^4x \cL \qty(\phi,\del_{\mu_1}\phi,\,\del_{\mu_1\mu_2} \phi,\ldots,\del_{\mu_1\cdots\mu_k}\phi,\dots)\,,
\end{align}
where $\phi(x)$ can in principle be any type of tensor field although we will eventually apply the result to our real scalar field theory; we denoted $\del_{\mu_1}\del_{\mu_2}\cdots\del_{\mu_k}$ by $\del_{\mu_1 \mu_2 \cdots \mu_k}$ where $k\in\mathbb{N}$ can be either finite or infinite.
The field equation reads
\begin{align}
&0=\frac{\del\cL}{\del\phi}
	- \del_{\nu_1} \frac{\del\cL}{\del(\del_{\nu_1}\phi)}
	\cdots
	+(-1)^{k} \del_{\nu_1 \nu_2 \cdots \nu_k}
	    \left\{\frac{\del\cL}{\del(\del_{\nu_1 \nu_2 \cdots \nu_k}\phi)}\right\}
	+\cdots \nonumber \\[2mm]
\Rightarrow&\quad
    \frac{\del\cL}{\del\phi}
    =-\sum_{k=1}^{N}
    (-1)^{k} \del_{\nu_1 \nu_2 \cdots \nu_k}
        \left\{ \frac{\del\cL}{\del(\del_{\nu_1 \nu_2 \cdots \nu_k}\phi)} \right\}\,,
\label{eq:the equation of the motion for generic high-del scalar}
\end{align}
where $N$ counts the number of derivatives acting on the field in the Lagrangian. Note that the variation with respect to the derivatives of the field is given by
\begin{align}
	\frac{\del(\del_{\mu_1\mu_2\ldots\mu_k}\phi)}{\del(\del_{\nu_1\nu_2\ldots\nu_k}\phi)}
		= \delta_{\mu_1}^{\nu_1} \delta_{\mu_2}^{\nu_2} \cdots \delta_{\mu_k}^{\nu_k}\,,
\end{align}
from which it is clear that, while $\del_{\mu_1\mu_2\ldots\mu_k}\phi$ is symmetric in all its indexes, the derivative $\del/\del(\del_{\mu_1\mu_2\ldots\mu_k}\phi)$ is not. We have to carefully take into account this fact when deriving the expression for the energy-momentum tensor.

Let us compute the derivative of the Lagrangian
\begin{align}
&\del_\alpha\cL
	=\frac{\del\cL}{\del\phi} \del_\alpha\phi
	    + \frac{\del\cL}{\del(\del_{\nu_1}\phi)} \del_\alpha\del_{\nu_1}\phi
		+ \frac{\del\cL}{\del(\del_{\nu_1\nu_2}\phi)} \del_\alpha\del_{\nu_1\nu_2}\phi
		+ \cdots
		+ \frac{\del\cL}{\del(\del_{\nu_1\nu_2\ldots\nu_k}\phi)} \del_\alpha\del_{\nu_1\nu_2\ldots\nu_k}\phi
		+ \cdots \nonumber \\
&\Rightarrow\quad
    \del_\alpha\cL
    =   \frac{\del\cL}{\del\phi} \del_\alpha\phi
        +\sum_{k=1}^{N}
            \frac{\del\cL}{\del(\del_{\nu_1\nu_2\ldots\nu_k}\phi)}
            \del_\alpha\del_{\nu_1\nu_2\ldots\nu_k}\phi
\label{derivative of the Lagraigian}\,,
\end{align}
and by using~\eqref{eq:the equation of the motion for generic high-del scalar} for the first term in the r.h.s. of~\eqref{derivative of the Lagraigian} we can write
\begin{align}
	\del_\alpha \cL
		= \sum_{k=1}^{N}
			\qty[
				\frac{\del\cL}{\del(\del_{\nu_1\nu_2\ldots\nu_k}\phi)} \del_\alpha\del_{\nu_1\nu_2\ldots\nu_k}\phi
				-(-1)^k \del_{\nu_1\nu_2\ldots\nu_k} \left\{\frac{\del\cL}{\del(\del_{\nu_1\nu_2\ldots\nu_k}\phi)}\right\}
						\del_\alpha\phi
			]\,.
\label{eq:recasted derivative of Lagrangian}
\end{align}
We now recast the terms inside the square brackets in a more suitable form by using the Leibniz rule.

We rewrite the $k=1$ term as
\begin{align}
    \frac{\del\cL}{\del(\del_{\nu_1}\phi)}\del_{\nu_1}\del_\alpha\phi
    + \del_{\nu_1}\frac{\del\cL}{\del(\del_{\nu_1}\phi)}\del_\alpha\phi
    = \del_{\nu_1} \qty[\frac{\del\cL}{\del(\del_{\nu_1}\phi)}\del_\alpha\phi]\,,
\end{align}
which corresponds to the standard contribution in a two-derivative theory.

The $k=2$ term, instead, reads
\begin{align}
&\frac{\del\cL}{\del(\del_{\nu_1\nu_2}\phi)}\del_{\nu_1\nu_2}\del_\alpha\phi
    -\del_{\nu_1\nu_2}\frac{\del\cL}{\del(\del_{\nu_1\nu_2}\phi)}\del_\alpha\phi
\nonumber\\[2mm]
&\qquad\qquad \qquad=
    \frac{\del\cL}{\del(\del_{\nu_1\nu_2}\phi)}\del_{\nu_1\nu_2}\del_\alpha\phi
        -\del_{\nu_1\nu_2}\frac{\del\cL}{\del(\del_{\nu_2\nu_1}\phi)}\del_\alpha\phi
\nonumber\\[2mm]
&\qquad\qquad \qquad=
    \del_{\nu_1}\qty[
        \frac{\del\cL}{\del(\del_{\nu_1\nu_2}\phi)}\del_{\nu_2}\del_\alpha\phi
        -\del_{\nu_2}\frac{\del\cL}{\del(\del_{\nu_2\nu_1}\phi)}\del_\alpha\phi ]
\nonumber\\[2mm]
&\qquad\qquad\qquad\quad
    -\del_{\nu_1}\left\{\frac{\del\cL}{\del(\del_{\nu_1\nu_2}\phi)}\right\}\del_{\nu_2}\del_\alpha\phi
    +\del_{\nu_2}\left\{\frac{\del\cL}{\del(\del_{\nu_1\nu_2}\phi)}\right\}\del_{\nu_1}\del_\alpha\phi
\nonumber\\[2mm]
&\qquad\qquad\qquad=
    \del_{\nu_1}\qty[
        \frac{\del\cL}{\del(\del_{\nu_1\nu_2}\phi)}\del_{\nu_2}\del_\alpha\phi
        -\del_{\nu_2}\frac{\del\cL}{\del(\del_{\nu_2\nu_1}\phi)}\del_\alpha\phi ]\,,
\end{align}
where we have used the fact that $\del_{\nu_1\nu_2}\phi$ is symmetric and that $\nu_1$ and $\nu_2$ are dummy indexes.
It is not difficult to guess the form for a generic term with a generic $k$:
\begin{align}
&\frac{\del\cL}{\del(\del_{\nu_1\nu_2\ldots\nu_k}\phi)} \del_\alpha\del_{\nu_1\nu_2\ldots\nu_k}\phi
	-(-1)^k \del_{\nu_1\nu_2\ldots\nu_k}\left\{\frac{\del\cL}{\del(\del_{\nu_1\nu_2\ldots\nu_k}\phi)}\right\}
		\del_\alpha\phi
\nonumber\\[2mm]
&\qquad\qquad\qquad
=\del_{\nu_1}\left[
        \frac{\del\cL}{\del(\del_{\nu_1\nu_2\ldots\nu_k}\phi)} \del_\alpha\del_{\nu_2\nu_3\ldots\nu_k}\phi
        - \del_{\nu_2} \left\{\frac{\del\cL}{\del(\del_{\nu_2\nu_1\ldots\nu_k}\phi)}\right\}
            \del_\alpha\del_{\nu_3\nu_4\ldots\nu_k}\phi
\right.\nonumber\\[2mm]
&\qquad\qquad\qquad\quad
\left.
        + \del_{\nu_2\nu_3} \left\{\frac{\del\cL}{\del(\del_{\nu_2\nu_3\nu_1\ldots\nu_k}\phi)}\right\}
            \del_\alpha\del_{\nu_4\nu_5\ldots\nu_k}\phi+\cdots
\right.\nonumber\\[2mm]
&\qquad\qquad\qquad\qquad\left.
        \cdots+ (-1)^{k+1}
            \del_{\nu_2\cdots\nu_k} \left\{\frac{\del\cL}{\del(\del_{\nu_2\nu_3\ldots\nu_k\nu_1}\phi)}\right\}
            \del_\alpha\phi
    \right]\,.
\end{align}
The above expressions at each $k$-th order were also derived in Ref.~\cite{Moeller:2002vx}; see also Refs.~\cite{Heredia:2021wja,Heredia:2022mls} for complementary works on the energy-momentum tensor in nonlocal field theories. In what follows, we will rewrite those expressions in a compact form by using some properties of functional derivatives and summations.

First of all, we can recast the term inside the square brackets in Eq.~\eqref{eq:recasted derivative of Lagrangian} (including all $k$-th orders) as
\begin{align}
    \del_{\nu_1} \qty[
			\sum_{l=1}^k (-1)^{l+1}
			\del^{(l-1)}_{\nu_2\ldots\nu_l} \left\{\frac{\del\cL}{\del(\del_{\nu_2\ldots\nu_l\nu_1\nu_{l+1}\ldots\nu_k}\phi)}\right\}
			\del^{(k-l)}_{\nu_{l+1}\ldots\nu_k} \del_\alpha\phi
		]\,,
\end{align}
where for $l=1$ the indexes $\nu_2\ldots\nu_l\nu_1\nu_{l+1}\ldots\nu_k$ should be understood as $\nu_1\nu_2\ldots\nu_k$.

We can rewrite the derivative of the Lagrangian~\eqref{eq:recasted derivative of Lagrangian} as a  total derivative,
\begin{align}
&	\del_\alpha \cL
		= \del_{\nu_1} \qty[
				\sum_{k=1}^{N} \sum_{l=1}^k (-1)^{l+1}
				\del^{(l-1)}_{\nu_2\ldots\nu_l} \left\{\frac{\del\cL}{\del(\del_{\nu_2\ldots\nu_l\nu_1\nu_{l+1}\ldots\nu_k}\phi)}\right\}
				\del^{(k-l)}_{\nu_{l+1}\ldots\nu_k} \del_\alpha\phi
		] \nonumber \\[2mm]
&	\Rightarrow\quad
	\del_{\nu_1} \qty[
				\sum_{k=1}^{N} \sum_{l=1}^k (-1)^{l+1}
					\del^{(l-1)}_{\nu_2\ldots\nu_l} \left\{\frac{\del\cL}{\del(\del_{\nu_2\ldots\nu_l\nu_1\nu_{l+1}\ldots\nu_k}\phi)}\right\}
						\del^{(k-l)}_{\nu_{l+1}\ldots\nu_k} \del_\alpha\phi
				-\delta^{\nu_1}_{\alpha} \cL
			]=0\,.
\end{align}
Finally, we can extract the conserved energy-momentum tensor $T^{\nu_1}_{\alpha}$ which is defined as
\begin{align}
T^{\nu_1}_{\alpha}
		=\sum_{k=1}^{N} \sum_{l=1}^k (-1)^{l+1}
				\del^{(l-1)}_{\nu_2\ldots\nu_l} \left\{\frac{\del\cL}{\del(\del_{\nu_2\ldots\nu_l\nu_1\nu_{l+1}\ldots\nu_k}\phi)}\right\}
					\del^{(k-l)}_{\nu_{l+1}\ldots\nu_k} \del_\alpha\phi
			-\delta^{\nu_1}_{\alpha} \cL\,,
\label{eq:generic energy-momentum tensor in higher-derivative theories}
\end{align}
and it is indeed conserved as it satisfies the conservation law $\del_{\nu_1}T^{\nu_1}_{\alpha}=0$.

\subsection{Nonlocal real scalar field}
We now apply the formula~\eqref{eq:generic energy-momentum tensor in higher-derivative theories} to the case of a nonlocal real scalar field $\phi$ with a  mass $m$ and a potential term $V(\phi)$ in $1+3$ dimensions. In particular, we consider the action
\begin{align}
    \cL=
        \frac{1}{2} \phi F(-\Box) \qty(\Box-m^2) \phi - V(\phi)\,,
        \label{eq:nonlocal scalar field action in app A}
\end{align}
where the potential term $V(\phi)$ contains cubic and higher powers of $\phi$, and the differential operator can be expressed as
\begin{align}
F(-\Box)=\sum\limits_{n=0}^\infty f_n(-1)^n\Box^n \,;
\label{eq:form-factor}
\end{align}
in this case $N=\infty.$
For convenience, we denote the kinetic term and the mass term by $\cL_K$ and $\cL_m,$ i.e.
\begin{align}
    \cL_K &= \frac{1}{2} \phi F(-\Box) \Box \phi\,, \\
    \cL_m &= - \frac{1}{2} m^2 \phi F(-\Box) \phi\,.
\end{align}
We now compute~\eqref{eq:generic energy-momentum tensor in higher-derivative theories} for the above Lagrangian.
By taking the derivatives of $\cL_K$ we have:
\begin{align}
\frac{\del\cL_K}{\del(\del_{\nu_2\ldots\nu_l\nu_1\nu_{l+1}\ldots\nu_k}\phi)}
&
    =\frac{1}{2} \sum_{n=0}^{\infty} f_n\qty(-1)^n
        \frac{\del}{\del(\del_{\nu_2\ldots\nu_l\nu_1\nu_{l+1}\ldots\nu_k}\phi)}
        \qty( \phi \Box^{n+1} \phi ) \nonumber\\[2mm]
&
    =\frac{1}{2} \phi \sum_{n=0}^{\infty} f_n \qty(-1)^n
        \frac{\del}{\del(\del_{\nu_2\ldots\nu_l\nu_1\nu_{l+1}\ldots\nu_k}\phi)}
        \qty( \qty(\del_\alpha \del^\alpha)^{n+1} \phi ) \nonumber\\[2mm]
&
    =\frac{1}{2} \phi \sum_{n=0}^{\infty} f_n \qty(-1)^n
        \eta^{\alpha_1\beta_1}\cdots\eta^{\alpha_{n+1}\beta_{n+1}}
        \frac{\del\qty(\del_{\alpha_1\beta_1\cdots\alpha_{n+1}\beta_{n+1}} \phi)}
                {\del\qty(\del_{\nu_2\ldots\nu_l\nu_1\nu_{l+1}\ldots\nu_k}\phi)}\,.
\end{align}
If $k=2(n+1)$ the derivative is  non-zero and we get
\begin{align}
\frac{\del\qty(\del_{\alpha_1\beta_1\cdots\alpha_{n+1}\beta_{n+1}} \phi)}
        {\del\qty(\del_{\nu_2\ldots\nu_l\nu_1\nu_{l+1}\ldots\nu_k}\phi)}
    =\delta_{\alpha_1\beta_1\cdots\alpha_{n+1}\beta_{n+1}}^{\nu_2\ldots\nu_l\nu_1\nu_{l+1}\ldots\nu_k} \delta_{k,2(n+1)}\,,
\end{align}
where we used the notation $\delta_{\nu_1}^{\mu_1}\delta_{\nu_2}^{\mu_2}\cdots\delta_{\nu_n}^{\mu_n}=\delta_{\nu_1 \nu_2 \cdots \nu_n}^{\mu_1 \mu_2 \cdots \mu_n}.$
Thus, we obtain
\begin{align}
\frac{\del\cL_K}{\del(\del_{\nu_2\ldots\nu_l\nu_1\nu_{l+1}\ldots\nu_k}\phi)}
    =\frac{1}{2} \phi \sum_{n=0}^{\infty}f_n \qty(-1)^n
        \eta^{\alpha_1\beta_1}\cdots\eta^{\alpha_{n+1}\beta_{n+1}}
        \delta_{\alpha_1\beta_1\cdots\alpha_{n+1}\beta_{n+1}}^{\nu_2\ldots\nu_l\nu_1\nu_{l+1}\ldots\nu_k} \delta_{k,2(n+1)}\,.
\end{align}
Similarly we can compute the mass-term as
\begin{align}
\frac{\del\cL_m}{\del(\del_{\nu_2\ldots\nu_l\nu_1\nu_{l+1}\ldots\nu_k}\phi)}
&
    = - \frac{1}{2} m^2 \sum_{n=0}^{\infty} f_n \qty(-1)^n
        \frac{\del}{\del(\del_{\nu_2\ldots\nu_l\nu_1\nu_{l+1}\ldots\nu_k}\phi)}
        \qty( \phi \Box^{n} \phi )\nonumber \\[2mm]
&
    = - \frac{1}{2} m^2 \phi \sum_{n=0}^{\infty}f_n \qty(-1)^n
        \frac{\del}{\del(\del_{\nu_2\ldots\nu_l\nu_1\nu_{l+1}\ldots\nu_k}\phi)}
        \qty( \qty(\del_\alpha \del^\alpha)^{n} \phi )\nonumber \\[2mm]
&
    = - \frac{1}{2} m^2 \phi \sum_{n=0}^{\infty} f_n \qty(-1)^n
        \eta^{\alpha_1\beta_1}\cdots\eta^{\alpha_{n}\beta_{n}}
        \frac{\del\qty(\del_{\alpha_1\beta_1\cdots\alpha_{n}\beta_{n}} \phi)}
                {\del\qty(\del_{\nu_2\ldots\nu_l\nu_1\nu_{l+1}\ldots\nu_k}\phi)} \nonumber\\[2mm]
&
    = - \frac{1}{2} m^2 \phi \sum_{n=0}^{\infty} f_n \qty(-1)^n
        \eta^{\alpha_1\beta_1}\cdots\eta^{\alpha_{n}\beta_{n}}
        \delta_{\alpha_1\beta_1\cdots\alpha_{n}\beta_{n}}^{\nu_2\ldots\nu_l\nu_1\nu_{l+1}\ldots\nu_k} \delta_{k,2n}\,.
\end{align}
Substituting the above expressions for the derivatives of the Lagrangian with respect to the field's derivatives into~\eqref{eq:generic energy-momentum tensor in higher-derivative theories}, we obtain
\begin{align}
T^{\nu_1}_{\alpha}&=
	\frac{1}{2} \eta^{\alpha_1\beta_1}\cdots\eta^{\alpha_{n+1}\beta_{n+1}}
	\sum_{k=1}^{\infty} \sum_{l=1}^k\sum_{n=0}^{\infty}
	f_n\qty(-1)^n(-1)^{l+1} 
	\delta_{\alpha_1\beta_1\cdots\alpha_{n+1}\beta_{n+1}}^{\nu_2\ldots\nu_l\nu_1\nu_{l+1}\ldots\nu_k} \delta_{k,2(n+1)}
\nonumber\\[2mm]
&\qquad\qquad\qquad\qquad\qquad\qquad\qquad\qquad\times
		\qty(\del^{(l-1)}_{\nu_2\ldots\nu_l} \phi)
		\qty(\del^{(k-l)}_{\nu_{l+1}\ldots\nu_k} \del_\alpha\phi)
\nonumber\\[2mm]
&\qquad
	 - \frac{1}{2} m^2 \eta^{\alpha_1\beta_1}\cdots\eta^{\alpha_{n}\beta_{n}}
	\sum_{k=1}^{\infty} \sum_{l=1}^k\sum_{n=0}^{\infty}
	f_n\qty(-1)^n(-1)^{l+1} 
	\delta_{\alpha_1\beta_1\cdots\alpha_{n}\beta_{n}}^{\nu_2\ldots\nu_l\nu_1\nu_{l+1}\ldots\nu_k} \delta_{k,2n}
\nonumber\\[2mm]
&\qquad\qquad\qquad\qquad\qquad\qquad\qquad\qquad\qquad\times
		\qty(\del^{(l-1)}_{\nu_2\ldots\nu_l} \phi)
		\qty(\del^{(k-l)}_{\nu_{l+1}\ldots\nu_k} \del_\alpha\phi)
	-\delta^{\nu_1}_{\alpha} \cL\,.
	\label{eq:stress-energ-delta}
\end{align}
%
%
Let us introduce the new index $m$ instead of $k$ through the change $k=l+m,$ so that given a function $f(n,l,k)$ we can write
\begin{align}
\sum_{k=1}^{\infty} \sum_{l=1}^k \sum_{n=0}^{\infty}
        f(n,l,k) \delta_{k,2(n+1)}
    &=\sum_{m=0}^{\infty} \sum_{l=1}^\infty \sum_{n=0}^{\infty}
            f(n,l,l+m) \delta_{l+m,2(n+1)} \nonumber \\[2mm]
    &=\sum_{n=0}^{\infty} \sum_{m=0}^{2n+1}
            f(n,2(n+1)-m,2(n+1))\,,
\end{align}
where the Kronecker delta was used to remove the $l$-summation in second line through the condition $m=2(n+1)-l \Leftrightarrow l=2(n+1)-m;$ the $m$-summation now runs over $0\leq m \leq 2n+1$ because in the $l$-summation we had  $1\leq l \leq k=2(n+1)$. The same holds for the other Kronecker delta $\delta_{k,2n},$ and in this case $2n = l+m \geq 1$ so that the $n$-summation runs over $n\geq 1$:
\begin{align}
\sum_{k=1}^{\infty} \sum_{l=1}^k \sum_{n=0}^{\infty}
        f(n,l,k) \delta_{k,2n}
    &=\sum_{m=0}^{\infty} \sum_{l=1}^\infty \sum_{n=0}^{\infty}
            f(n,l,l+m) \delta_{l+m,2n} \nonumber \\[2mm]
    &=\sum_{n=1}^{\infty} \sum_{m=0}^{2n-1}
            f(n,2n-m,2n)\,.
\end{align}
By applying the above two formula to the terms in the expression of the energy-momentum tensor~\eqref{eq:stress-energ-delta}, we obtain
\begin{align}
&\sum_{k=1}^{\infty} \sum_{l=1}^k \sum_{n=0}^{\infty}
	f_n\qty(-1)^n(-1)^{l+1} 
	\delta_{\alpha_1\beta_1\cdots\alpha_{n+1}\beta_{n+1}}^{\nu_2\ldots\nu_l\nu_1\nu_{l+1}\ldots\nu_k} \delta_{k,2(n+1)}
		\qty(\del^{(l-1)}_{\nu_2\ldots\nu_l} \phi)
		\qty(\del^{(k-l)}_{\nu_{l+1}\ldots\nu_k} \del_\alpha\phi) \nonumber\\[2mm]
&=\sum_{n=0}^{\infty} \sum_{m=0}^{2n+1}
	f_n\qty(-1)^n(-1)^{m+1}
	\delta
	    _{\alpha_1\beta_1\cdots\alpha_{n+1}\beta_{n+1}}
	    ^{\nu_2\ldots\nu_{2(n+1)-m}\nu_1\nu_{2(n+1)-m+1}\ldots\nu_{2(n+1)}}
\nonumber\\&\qquad\qquad\qquad\qquad\qquad\qquad\times
	\qty(\del^{(2(n+1)-m-1)}_{\nu_2\ldots\nu_{2(n+1)-m}} \phi)
	\qty(\del^{(m)}_{\nu_{2(n+1)-m+1}\ldots\nu_{2(n+1)}} \del_\alpha\phi)\,,
\end{align}
and
\begin{align}
&\sum_{k=1}^{\infty} \sum_{l=1}^k \sum_{n=1}^{\infty}
	f_n \qty(-1)^n(-1)^{l+1} 
	\delta_{\alpha_1\beta_1\cdots\alpha_{n}\beta_{n}}^{\nu_2\ldots\nu_l\nu_1\nu_{l+1}\ldots\nu_k} \delta_{k,2n}
		\qty(\del^{(l-1)}_{\nu_2\ldots\nu_l} \phi)
		\qty(\del^{(k-l)}_{\nu_{l+1}\ldots\nu_k} \del_\alpha\phi) \nonumber\\[2mm]
&=\sum_{n=1}^{\infty} \sum_{m=0}^{2n-1}
	f_n \qty(-1)^n(-1)^{m+1}
	\delta
	    _{\alpha_1\beta_1\cdots\alpha_{n}\beta_{n}}
	    ^{\nu_2\ldots\nu_{2n-m}\nu_1\nu_{2n-m+1}\ldots\nu_{2n}}
	\qty(\del^{(2n-m-1)}_{\nu_2\ldots\nu_{2n-m}} \phi)
	\qty(\del^{(m)}_{\nu_{2n-m+1}\ldots\nu_{2n}} \del_\alpha\phi)\,.
\end{align}
Then $T^{\nu_1}_\alpha$ can be recast as
\begin{align}
T^{\nu_1}_{\alpha}&=
	\frac{1}{2} \eta^{\alpha_1\beta_1}\cdots\eta^{\alpha_{n+1}\beta_{n+1}}
	\sum_{n=0}^{\infty} f_n \qty(-1)^n
	\sum_{m=1}^{2n+1} (-1)^{m+1}
	\delta
	    _{\alpha_1\beta_1\cdots\alpha_{n+1}\beta_{n+1}}
	    ^{\nu_2\ldots\nu_{2(n+1)-m}\nu_1\nu_{2(n+1)-m+1}\ldots\nu_{2(n+1)}}
\nonumber\\[2mm]
&\quad\qquad\qquad\qquad\qquad\qquad\times
	\qty(\del^{(2(n+1)-m-1)}_{\nu_2\ldots\nu_{2(n+1)-m}} \phi)
	\qty(\del^{(m)}_{\nu_{2(n+1)-m+1}\ldots\nu_{2(n+1)}} \del_\alpha\phi)
\nonumber\\[2mm]
&\quad-
	\frac{1}{2} m^2 \, \eta^{\alpha_1\beta_1}\cdots\eta^{\alpha_{n}\beta_{n}}
	\sum_{n=1}^{\infty} f_n\qty(-1)^n
	\sum_{m=1}^{2n-1} (-1)^{m+1}
	\delta
	    _{\alpha_1\beta_1\cdots\alpha_{n}\beta_{n}}
	    ^{\nu_2\ldots\nu_{2n-m}\nu_1\nu_{2n-m+1}\ldots\nu_{2n}}
\nonumber\\[2mm]
&\quad\qquad\qquad\qquad\qquad\qquad\qquad\times
	\qty(\del^{(2n-m-1)}_{\nu_2\ldots\nu_{2n-m}} \phi)
	\qty(\del^{(m)}_{\nu_{2n-m+1}\ldots\nu_{2n}} \del_\alpha\phi)
	-\delta^{\nu_1}_{\alpha} \cL\,.
\end{align}
Next we have to deal with the Kronecker deltas in the first and the second terms.
To do so, we decompose the $m$-summation into the even  $m=2s$ and odd $m=2s+1$ parts, with the $s$-summation running over $0\leq s \leq n$. Then, the first term becomes
\begin{align}
&
\sum_{m=0}^{2n+1} (-1)^{m+1}
	\delta
	    _{\alpha_1\beta_1\cdots\alpha_{n+1}\beta_{n+1}}
	    ^{\nu_2\ldots\nu_{2(n+1)-m}\nu_1\nu_{2(n+1)-m+1}\ldots\nu_{2(n+1)}}
	\qty(\del^{(2(n+1)-m-1)}_{\nu_2\ldots\nu_{2(n+1)-m}} \phi)
	\qty(\del^{(m)}_{\nu_{2(n+1)-m+1}\ldots\nu_{2(n+1)}} \del_\alpha\phi)
	\nonumber\\[2mm]
&=
\sum_{s=0}^{n}\left[
	-\delta
	    _{\alpha_1\beta_1\cdots\alpha_{n+1}\beta_{n+1}}
	    ^{\nu_2\ldots\nu_{2(n+1-s)}\nu_1\nu_{2(n+1-s)+1}\ldots\nu_{2(n+1)}}
	\qty(\del^{(2(n+1-s)-1)}_{\nu_2\ldots\nu_{2(n+1-s)}} \phi)
	\qty(\del^{(2s)}_{\nu_{2(n+1-s)+1}\ldots\nu_{2(n+1)}} \del_\alpha\phi)
\right.\nonumber\\[2mm]
&\qquad\qquad+\left.
	\delta
	    _{\alpha_1\beta_1\cdots\alpha_{n+1}\beta_{n+1}}
	    ^{\nu_2\ldots\nu_{2(n+1-s)-1}\nu_1\nu_{2(n+1-s)}\ldots\nu_{2(n+1)}}
	\qty(\del^{(2(n+1-s))}_{\nu_2\ldots\nu_{2(n+1-s)-1}} \phi)
	\qty(\del^{(2s+1)}_{\nu_{2(n+1-s)}\ldots\nu_{2(n+1)}} \del_\alpha\phi)
	\right]
	\nonumber\\[2mm]
&=
\sum_{s=0}^{n}\left[
	-\delta_{\alpha_1}^{\nu_2}
	    \delta_{\beta_1}^{\nu_3}
	    \cdots
	        \delta_{\alpha_{n+1-s}}^{\nu_{2(n+1-s)}}
	        \delta_{\beta_{n+1-s}}^{\nu_1}
	        \delta_{\alpha_{n+2-s}}^{\nu_{2(n+1-s)+1}}
	    \cdots
	        \delta_{\alpha_{n+1}}^{\nu_{2n+1}}
	        \delta_{\beta_{n+1}}^{\nu_{2(n+1)}}
\right.\nonumber\\[2mm]
&\qquad\qquad\qquad\qquad\qquad\qquad\times
	\qty(\del^{(2(n+1-s)-1)}_{\nu_2\ldots\nu_{2(n+1-s)}} \phi)
	\qty(\del^{(2s)}_{\nu_{2(n+1-s)+1}\ldots\nu_{2(n+1)}} \del_\alpha\phi)
\nonumber\\[2mm]
&\qquad\qquad+
	\delta_{\alpha_1}^{\nu_2}
	    \delta_{\beta_1}^{\nu_3}
	    \cdots
	        \delta_{\beta_{n-s}}^{\nu_{2(n+1-s)-1}}
	        \delta_{\alpha_{n-s+1}}^{\nu_1}
	        \delta_{\beta_{n-s+1}}^{\nu_{2(n+1-s)}}
	    \cdots
	        \delta_{\alpha_{n+1}}^{\nu_{2n+1}}
	        \delta_{\beta_{n+1}}^{\nu_{2(n+1)}}
\nonumber\\[2mm]
&\qquad\qquad\qquad\qquad\qquad\qquad\qquad\times\left.
	\qty(\del^{(2(n+1-s))}_{\nu_2\ldots\nu_{2(n+1-s)-1}} \phi)
	\qty(\del^{(2s+1)}_{\nu_{2(n+1-s)}\ldots\nu_{2(n+1)}} \del_\alpha\phi)
	\right]
	\nonumber\\[2mm]
&=
\sum_{s=0}^{n}
    \delta_{\alpha_1}^{\nu_2} \delta_{\beta_1}^{\nu_3}
    \cdots
        \delta_{\alpha_{n-s}}^{\nu_{2(n-s)}}
        \delta_{\beta_{n-s}}^{\nu_{2(n-s)+1}}
        \delta_{\alpha_{n-s+2}}^{\nu_{2(n+1-s)+1}}
    \cdots
        \delta_{\alpha_{n+1}}^{\nu_{2n+1}}
        \delta_{\beta_{n+1}}^{\nu_{2(n+1)}}
\nonumber\\[2mm]
&\qquad\qquad\times\left[
    -\delta_{\alpha_{n-s+1}}^{\nu_{2(n+1-s)}} \delta_{\beta_{n-s+1}}^{\nu_1}
        \qty(\del^{(2(n+1-s)-1)}_{\nu_2\ldots\nu_{2(n+1-s)}} \phi)
        \qty(\del^{(2s)}_{\nu_{2(n+1-s)+1}\ldots\nu_{2(n+1)}} \del_\alpha\phi)
\right.\nonumber\\[2mm]
&\qquad\qquad\qquad\qquad\left.
    +\delta_{\alpha_{n-s+1}}^{\nu_1} \delta_{\beta_{n-s+1}}^{\nu_{2(n+1-s)}}
	    \qty(\del^{(2(n+1-s))}_{\nu_2\ldots\nu_{2(n+1-s)-1}} \phi)
	    \qty(\del^{(2s+1)}_{\nu_{2(n+1-s)}\ldots\nu_{2(n+1)}} \del_\alpha\phi)
\right]\nonumber\\[2mm]
&\longrightarrow\,\,\sum_{s=0}^{n}
    \delta_{\alpha_1}^{\nu_2} \delta_{\beta_1}^{\nu_3}
    \cdots
        \delta_{\alpha_{n-s}}^{\nu_{2(n-s)}}
        \delta_{\beta_{n-s}}^{\nu_{2(n-s)+1}}
        \delta_{\alpha_{n-s+1}}^{\nu_{2(n+1-s)}}
        \delta_{\beta_{n-s+1}}^{\nu_1}
        \delta_{\alpha_{n-s+2}}^{\nu_{2(n+1-s)+1}}
    \cdots
        \delta_{\alpha_{n+1}}^{\nu_{2n+1}}
        \delta_{\beta_{n+1}}^{\nu_{2(n+1)}}
\nonumber\\[2mm]
&\qquad\qquad\times\left[
    -\qty(\del^{(2(n+1-s)-1)}_{\nu_2\ldots\nu_{2(n+1-s)}} \phi)
        \qty(\del^{(2s)}_{\nu_{2(n+1-s)+1}\ldots\nu_{2(n+1)}} \del_\alpha\phi)
\right.\nonumber\\[2mm]
&\qquad\qquad\qquad\qquad\qquad\qquad\left.
    +\qty(\del^{(2(n+1-s))}_{\nu_2\ldots\nu_{2(n+1-s)-1}} \phi)
	    \qty(\del^{(2s+1)}_{\nu_{2(n+1-s)}\ldots\nu_{2(n+1)}} \del_\alpha\phi)
\right]\,;
\end{align}
in the last step, the arrow means that the dummy indexes $\alpha_{n-s+1}$ and $\beta_{n-s+1}$ have been exchanged as in the expression of the energy-momentum tensor one can use the symmetry property of the metric tensor $\eta^{\alpha_{n-s+1}\beta_{n-s+1}}$. 

An analogous form can be obtained for the mass-term. In this case $0\leq s \leq n-1$ and we get
\begin{align}
&
\sum_{m=0}^{2n-1} (-1)^{m+1}
	\delta
	    _{\alpha_1\beta_1\cdots\alpha_{n}\beta_{n}}
	    ^{\nu_2\ldots\nu_{2n-m}\nu_1\nu_{2n-m+1}\ldots\nu_{2n}}
	\qty(\del^{(2n-m-1)}_{\nu_2\ldots\nu_{2n-m}} \phi)
	\qty(\del^{(m)}_{\nu_{2n-m+1}\ldots\nu_{2n}} \del_\alpha\phi)
	\nonumber\\[2mm]
&=
\sum_{s=0}^{n-1}\left[
	-\delta_{\alpha_1}^{\nu_2}
	    \delta_{\beta_1}^{\nu_3}
	    \cdots
	        \delta_{\alpha_{n-s}}^{\nu_{2(n-s)}}
	        \delta_{\beta_{n-s}}^{\nu_1}
	        \delta_{\alpha_{n+1-s}}^{\nu_{2(n-s)+1}}
	    \cdots
	        \delta_{\alpha_{n}}^{\nu_{2n-1}}
	        \delta_{\beta_{n}}^{\nu_{2n}}
\right. 
	\qty(\del^{(2(n-s)-1)}_{\nu_2\ldots\nu_{2(n-s)}} \phi)
	\qty(\del^{(2s)}_{\nu_{2(n-s)+1}\ldots\nu_{2n}} \del_\alpha\phi)
\nonumber\\[2mm]
&\qquad\qquad+
	\delta_{\alpha_1}^{\nu_2}
	    \delta_{\beta_1}^{\nu_3}
	    \cdots
	        \delta_{\beta_{n-1-s}}^{\nu_{2(n-s)-1}}
	        \delta_{\alpha_{n-s}}^{\nu_1}
	        \delta_{\beta_{n-s}}^{\nu_{2(n-s)}}
	    \cdots
	        \delta_{\alpha_{n}}^{\nu_{2n-1}}
	        \delta_{\beta_{n}}^{\nu_{2n}}
\left.
	\qty(\del^{(2(n-s))}_{\nu_2\ldots\nu_{2(n-s)-1}} \phi)
	\qty(\del^{(2s+1)}_{\nu_{2(n-s)}\ldots\nu_{2n}} \del_\alpha\phi)
	\right]
	\nonumber\\[2mm]
&\longrightarrow\,\,
\sum_{s=0}^{n-1}
    \delta_{\alpha_1}^{\nu_2} \delta_{\beta_1}^{\nu_3}
    \cdots
        \delta_{\alpha_{n-1-s}}^{\nu_{2(n-1-s)}}
        \delta_{\beta_{n-1-s}}^{\nu_{2(n-1-s)+1}}
        \delta_{\alpha_{n-s}}^{\nu_{2(n-s)}}
        \delta_{\beta_{n-s}}^{\nu_1}
        \delta_{\alpha_{n-s+1}}^{\nu_{2(n-s)+1}}
    \cdots
        \delta_{\alpha_{n}}^{\nu_{2n-1}}
        \delta_{\beta_{n}}^{\nu_{2n}}
\nonumber\\[2mm]
&\qquad\quad\times\left[
    -\qty(\del^{(2(n-s)-1)}_{\nu_2\ldots\nu_{2(n-s)}} \phi)
        \qty(\del^{(2s)}_{\nu_{2(n-s)+1}\ldots\nu_{2n}} \del_\alpha\phi)
    +\qty(\del^{(2(n-s))}_{\nu_2\ldots\nu_{2(n-s)-1}} \phi)
	    \qty(\del^{(2s+1)}_{\nu_{2(n-s)}\ldots\nu_{2n}} \del_\alpha\phi)
\right]\,.
\end{align}
Substituting the above expressions into the energy-momentum tensor $T_\alpha^{\nu_1}$, we obtain
\begin{align}
T^{\nu_1}_{\alpha}&=
	\frac{1}{2} \eta^{\alpha_1\beta_1}\cdots\eta^{\alpha_{n+1}\beta_{n+1}}
	\sum_{n=0}^{\infty} f_n \qty(-1)^n
\nonumber\\[2mm]
&\qquad\times
    \sum_{s=0}^{n}
    \delta_{\alpha_1}^{\nu_2} \delta_{\beta_1}^{\nu_3}
    \cdots
        \delta_{\alpha_{n-s}}^{\nu_{2(n-s)}}
        \delta_{\beta_{n-s}}^{\nu_{2(n-s)+1}}
        \delta_{\alpha_{n-s+1}}^{\nu_{2(n+1-s)}}
        \delta_{\beta_{n-s+1}}^{\nu_1}
        \delta_{\alpha_{n-s+2}}^{\nu_{2(n+1-s)+1}}
    \cdots
        \delta_{\alpha_{n+1}}^{\nu_{2n+1}}
        \delta_{\beta_{n+1}}^{\nu_{2(n+1)}}
\nonumber\\[2mm]
&\qquad\qquad\times
    \left[
    -\qty(\del^{(2(n+1-s)-1)}_{\nu_2\ldots\nu_{2(n+1-s)}} \phi)
        \qty(\del^{(2s)}_{\nu_{2(n+1-s)+1}\ldots\nu_{2(n+1)}} \del_\alpha\phi)
\right.\nonumber\\[2mm]
&\qquad\qquad\qquad\qquad\qquad\qquad\qquad\qquad\qquad\left.
    +\qty(\del^{(2(n+1-s))}_{\nu_2\ldots\nu_{2(n+1-s)-1}} \phi)
	    \qty(\del^{(2s+1)}_{\nu_{2(n+1-s)}\ldots\nu_{2(n+1)}} \del_\alpha\phi)\right]
\nonumber\\[2mm]
&\quad-
	\frac{1}{2} m^2 \, \eta^{\alpha_1\beta_1}\cdots\eta^{\alpha_{n}\beta_{n}}
	\sum_{n=1}^{\infty} f_n \qty(-1)^n
\nonumber\\[2mm]
&\qquad\quad\times
    \sum_{s=0}^{n-1}
        \delta_{\alpha_1}^{\nu_2} \delta_{\beta_1}^{\nu_3}
        \cdots
            \delta_{\alpha_{n-1-s}}^{\nu_{2(n-1-s)}}
            \delta_{\beta_{n-1-s}}^{\nu_{2(n-1-s)+1}}
            \delta_{\alpha_{n-s}}^{\nu_{2(n-s)}}
            \delta_{\beta_{n-s}}^{\nu_1}
            \delta_{\alpha_{n-s+1}}^{\nu_{2(n-s)+1}}
        \cdots
            \delta_{\alpha_{n}}^{\nu_{2n-1}}
            \delta_{\beta_{n}}^{\nu_{2n}}
\nonumber\\[2mm]
&\qquad\qquad\quad\times\left[
        -\qty(\del^{(2(n-s)-1)}_{\nu_2\ldots\nu_{2(n-s)}} \phi)
            \qty(\del^{(2s)}_{\nu_{2(n-s)+1}\ldots\nu_{2n}} \del_\alpha\phi)
        +\qty(\del^{(2(n-s))}_{\nu_2\ldots\nu_{2(n-s)-1}} \phi)
    	    \qty(\del^{(2s+1)}_{\nu_{2(n-s)}\ldots\nu_{2n}} \del_\alpha\phi)
    \right]
\nonumber\\[2mm]
&\qquad
	-\delta^{\nu}_{\mu} \cL\,,
\label{eq:equation which will be refine by a decomposition to Beta and Gamma}
\end{align}
and making all possible contractions we get
\begin{align}
T_{\mu\nu}
&=
	\frac{1}{2} \sum_{n=0}^{\infty} f_n \qty(-1)^n
    \sum_{s=0}^{n}
    \left[
        \qty(\Box^{n-s}\phi) \qty(\Box^s\del_{\mu}\del_\nu\phi)
        -\qty(\Box^{n-s}\del_{\mu}\phi) \qty(\Box^s\del_\nu\phi)
    \right]
\nonumber\\[2mm]
&\quad-
	\frac{1}{2} m^2 \, \sum_{n=1}^{\infty} f_n \qty(-1)^n
    \sum_{s=0}^{n-1}
    \left[
        \qty(\Box^{n-s-1}\phi) \qty(\Box^s\del_{\mu}\del_\nu\phi)
        -\qty(\Box^{n-s-1}\del_{\mu}\phi) \qty(\Box^s\del_\nu\phi)
    \right]	-\eta_{\mu\nu} \cL\,,
\label{eq:conserved-T}
\end{align}
where we have replaced the indexes $\nu_1$ and $\alpha$ with $\nu$ and $\mu,$ respectively, and lowered both of them. One can verify that the energy-momentum tensor in Eq.~\eqref{eq:conserved-T} satisfy the conservation law $\partial^\mu T_{\mu\nu}=0.$

The expression~\eqref{eq:conserved-T} does not recover the standard form of the energy-momentum tensor in the two-derivative case because a total-derivative term should be added. Thus, we now add the following term in order to recover the appropriate two-derivative limit:
\begin{eqnarray}
W_{\mu\nu}&=&-\frac{1}{2}\sum\limits_{n=0}^\infty f_n (-1)^n\sum\limits_{s=0}^{n}\left[\partial_\mu(\Box^{n-s}\phi\Box^s\partial_\nu \phi)-\eta_{\mu\nu}\partial_\rho (\Box^s\phi\partial^\rho\Box^{n-s}\phi) \right] \nonumber\\[2mm]
&&+\frac{1}{2}m^2\sum\limits_{n=0}^\infty f_n (-1)^n\sum\limits_{s=0}^{n-1}\left[\partial_\mu(\Box^{n-1-s}\phi\Box^s\partial_\nu \phi)-\eta_{\mu\nu}\partial_\rho (\Box^s\phi\partial^\rho\Box^{n-1-s}\phi) \right]\,.
\end{eqnarray}
One can easily show that $\partial^\mu W_{\mu\nu}=0,$ which guarantees that the new energy-momentum tensor $T_{\mu\nu}\rightarrow T_{\mu\nu}+W_{\mu\nu}$ remains conserved.

%
%

The new expression of the energy-momentum tensor reads:
\begin{align}
T_{\mu\nu}
&=
-\eta_{\mu\nu} \mathcal{L}-\sum_{n=0}^{\infty} f_n (-1)^n
\sum_{s=0}^{n}
\left[\Box^s\partial_\mu\phi \,\Box^{n-s}\partial_\nu\phi
-\frac{1}{2}\eta_{\mu\nu}\partial_\rho\left(\Box^s\phi\,\partial^\rho\Box^{n-s}\phi\right)\right]
\nonumber\\[2mm]
&\!\!+
m^2 \, \sum_{n=1}^{\infty} f_n (-1)^n
\sum_{s=0}^{n-1}
\left[\Box^s\partial_\mu\phi \,\Box^{n-1-s}\partial_\nu\phi
-\frac{1}{2}\eta_{\mu\nu}\partial_\rho\left(\Box^s\phi\,\partial^\rho\Box^{n-1-s}\phi\right)\right]	\,.
\label{stress-final}
\end{align}
It is also worth to mention that the form of the energy-momentum tensor in Eq.~\eqref{stress-final} coincides with the Rosenfeld-Belinfante energy-momentum tensor that one would obtain by coupling the scalar field $\phi$ to gravity and taking the variation of the scalar field action with respect to the spacetime metric $g_{\mu\nu}.$

\bibliographystyle{utphys}
\bibliography{References}

\end{document}